\documentclass[11pt]{article}
\pdfoutput=1  
\usepackage{graphicx,color}
\usepackage{appendix}
\usepackage{latexsym,amsmath,amssymb,graphicx,booktabs}
\usepackage{epsfig,latexsym,cite}
\usepackage{hyperref}
\numberwithin{equation}{section}

\definecolor{MyBlue}{rgb}{0.15,0.15,0.70}

\hypersetup{
colorlinks=true,
citecolor=MyBlue,
linkcolor=MyBlue,
urlcolor=MyBlue
}

\input epsf
\usepackage{amssymb}
\usepackage{amsmath}
\usepackage{amsfonts}
\usepackage{upgreek}
\usepackage{latexsym}

\newcommand{\nn}{\nonumber}

\newcommand{\iBox}{\Box^{-1}}

\newcommand{\Fmn}{F_{\mu\nu}}
\newcommand{\FMN}{F^{\mu\nu}}

\renewcommand\({\left(}
\renewcommand\){\right)}
\renewcommand\[{\left[}
\renewcommand\]{\right]}

\newcommand\n{{\mbox {\boldmath $\nabla$}}}
\newcommand{\ra}{\rightarrow}

\def\lsim{\raise 0.4ex\hbox{$<$}\kern -0.8em\lower 0.62
ex\hbox{$\sim$}}

\def\gsim{\raise 0.4ex\hbox{$>$}\kern -0.7em\lower 0.62
ex\hbox{$\sim$}}

\def\lbar{{\hbox{$\lambda$}\kern -0.7em\raise 0.6ex
\hbox{$-$}}}

\newcommand\eq[1]{eq.~(\ref{#1})}
\newcommand\eqs[2]{eqs.~(\ref{#1}) and (\ref{#2})}
\newcommand\Eq[1]{Equation~(\ref{#1})}

\newcommand\eqss[3]{eqs.~(\ref{#1}), (\ref{#2}) and (\ref{#3})}
\newcommand\eqsss[4]{eqs.~(\ref{#1}), (\ref{#2}), (\ref{#3})
and (\ref{#4})}

\newcommand\eqst[2]{eqs.~(\ref{#1})--(\ref{#2})}

\newcommand\pa{\partial}
\newcommand\p{\partial}

\newcommand\ee{\end{equation}}
\newcommand\be{\begin{equation}}
\def\bea{\begin{array}}
\def\eea{\end{array}}\def\ea{\end{array}}
\newcommand\ees{\end{eqnarray}}
\newcommand\bees{\begin{eqnarray}}
\def\nn{\nonumber}

\def\a{\alpha}
\def\b{\beta}

\def\g{\gamma}

\def\d{\delta}

\def\dslash{\hspace{-1mm}\not{\hbox{\kern-2pt $\partial$}}}
\def\Dslash{\not{\hbox{\kern-2pt $D$}}}
\def\pslash{\not{\hbox{\kern-2.1pt $p$}}}
\def\kslash{\not{\hbox{\kern-2.3pt $k$}}}
\def\qslash{\not{\hbox{\kern-2.3pt $q$}}}

\newcommand{\vx}{{\bf x}}

\def\p1{{\bf p}_1}
\def\p2{{\bf p}_2}
\def\k1{{\bf k}_1}
\def\k2{{\bf k}_2}

\newcommand{\emn}{\eta_{\mu\nu}}

\newcommand{\gmn}{g_{\mu\nu}}

\newcommand{\gMN}{g^{\mu\nu}}
\newcommand{\gRS}{g^{\rho\sigma}}

\newcommand{\pam}{\pa_{\mu}}

\newcommand{\pan}{\pa_{\nu}}
\newcommand{\parho}{\pa_{\rho}}
\newcommand{\pas}{\pa_{\sigma}}
\newcommand{\paM}{\pa^{\mu}}

\newcommand{\paR}{\pa^{\rho}}

\newcommand{\Rmn}{R_{\mu\nu}}
\newcommand{\Gmn}{G_{\mu\nu}}
\newcommand{\RMN}{R^{\mu\nu}}

\newcommand{\Tmn}{T_{\mu\nu}}
\newcommand{\Smn}{S_{\mu\nu}}

\newcommand{\dddM}{\kern 0.2em \raise 1.9ex\hbox{$...$}\kern -1.0em \hbox{$M$}}
\newcommand{\dddQ}{\kern 0.2em \raise 1.9ex\hbox{$...$}\kern -1.0em \hbox{$Q$}}
\newcommand{\dddI}{\kern 0.2em \raise 1.9ex\hbox{$...$}\kern -1.0em\hbox{$I$}}
\newcommand{\dddJ}{\kern 0.2em \raise 1.9ex\hbox{$...$}\kern-1.0em
\hbox{$J$}}
\newcommand{\dddcalJ}{\kern 0.2em \raise 1.9ex\hbox{$...$}\kern-1.0em
\hbox{${\cal J}$}}

\newcommand{\dddO}{\kern 0.2em \raise 1.9ex\hbox{$...$}\kern -1.0em
\hbox{${\cal O}$}}
\def\dddz{\raise 1.5ex\hbox{$...$}\kern -0.8em \hbox{$z$}}
\def\dddd{\raise 1.8ex\hbox{$...$}\kern -0.8em \hbox{$d$}}
\def\dddbd{\raise 1.8ex\hbox{$...$}\kern -0.8em \hbox{${\bf d}$}}
\def\ddbd{\raise 1.8ex\hbox{$..$}\kern -0.8em \hbox{${\bf d}$}}
\def\dddx{\raise 1.6ex\hbox{$...$}\kern -0.8em \hbox{$x$}}

\newcommand{\Sch}{Schwarzschild }

\newcommand{\mplr}{m_{\rm Pl}}

\newcommand{\oma}{\Omega_{M}}
\newcommand{\ora}{\Omega_{R}}

\newcommand{\rde}{\rho_{\rm DE}}
\newcommand{\wde}{w_{\rm DE}}

\newcommand{\Uc}{U_{\rm cosmo}}

\DeclareMathOperator{\polylog}{Li_{\scriptscriptstyle2}}

\graphicspath{ {./Graphs/} }

\begin{document}

\begin{titlepage}

\vspace*{2cm}

\centerline{\Large \bf Testing nonlocal gravity with Lunar Laser Ranging\\}

\vspace{5mm}


\vskip 0.4cm
\vskip 0.7cm
\centerline{\large Enis Belgacem$^1$, Andreas Finke$^1$, Antonia Frassino$^{1,2}$ and  Michele Maggiore$^1$}
\vskip 0.3cm
\centerline{\em $^1$D\'epartement de Physique Th\'eorique and Center for Astroparticle Physics,}  
\centerline{\em Universit\'e de Gen\`eve, 24 quai Ansermet, CH--1211 Gen\`eve 4, Switzerland}
\vspace{3mm}
\centerline{\em $^2$Departament de F{\'\i}sica Qu\`antica i Astrof\'{\i}sica, Institut de
Ci\`encies del Cosmos,}
\centerline{\em  Universitat de
Barcelona, Mart\'{\i} i Franqu\`es 1, E-08028 Barcelona, Spain }  

\vskip 1.9cm

\begin{abstract}
We study the impact of the limit on $|\dot{G}|/G$ from Lunar Laser Ranging on  ``nonlocal gravity", i.e. on models of the quantum effective action of gravity that include  nonlocal terms relevant in the infrared, such as   the ``RR" and ``RT" models proposed by our group, and the Deser-Woodard (DW) model. We elaborate on  the analysis of Barreira {\em et al.} \cite{Barreira:2014kra} and we confirm their findings that (under plausible assumptions such as the absence of strong backreaction from non-linear structures), the RR model is  ruled out. We also show that the mechanism of ``perfect screening for free" suggested for the DW model actually does not work and  the DW model is also ruled out. In contrast, the RT model passes all  phenomenological consistency tests and is still a viable candidate.

\end{abstract}


\end{titlepage}

\newpage

\section{Introduction}

The study of infrared (IR) modifications of General Relativity  (GR) is largely motivated by  the aim of obtaining a dynamical understanding of dark energy (DE). One direction of investigation, that has been very much explored, is based on the addition of new degrees of freedom to GR as, for instance, in massive gravity, bigravity, or scalar-tensor theories. A different route consists in asking whether, already in GR, quantum effects could modify the IR behavior and give rise to interesting cosmological consequences. The idea that quantum gravity at large distances could induce cosmological effects is quite old, see e.g. Taylor and Veneziano~\cite{Taylor:1989ua}. A first-principle understanding of the IR structure of quantum gravity is a highly non-trivial task, but hints of non-trivial physics   in space-times of cosmological relevance emerge  for instance from the study of  their IR divergences~\cite{Tsamis:1994ca},  from the fact that
strong  IR effects appear for the conformal mode  \cite{Antoniadis:1986sb, Antoniadis:1991fa}, and from arguments against the quantum stability of de~Sitter space~\cite{Polyakov:2007mm,Polyakov:2012uc,Anderson:2013ila,Rajaraman:2016nvv,Dvali:2017eba} (see also recent discussions in the context of the ``swampland''~\cite{Ooguri:2006in,Agrawal:2018own,Dvali:2018jhn}).

When studying quantum effects, the relevant quantity is the quantum effective action, rather than the original fundamental action of the theory.  Working in terms of the quantum effective action opens new phenomenological possibilities, related to the introduction of nonlocal terms. Indeed,  according to the basic principles of QFT, the fundamental action of a theory must be local.  However, when in the spectrum there are  massless  particles (such as the graviton in GR), or more generally particles light with respect to the relevant energy scales,
the  quantum effective action  has also nonlocal terms. In particular, terms involving  inverse powers of the d'Alembertian are relevant in the IR and can affect the behavior of the theory at cosmological  scales. Given that a first-principle computation of the IR limit of the quantum effective action of gravity is quite difficult, especially if one is looking for non-perturbative effects, it makes sense to begin by adopting a phenomenological point of view and study what sort of nonlocal terms could give rise to a viable theory of dark energy.

In this spirit, a first nonlocal gravity model was proposed by Wetterich~\cite{Wetterich:1997bz}, who considered a quantum effective action of the form
\be\label{GammaWett}
\Gamma=\frac{\mplr^2}{2}\int d^4x \sqrt{-g}\,\[ R-g^2 R\iBox R\]\, ,
\ee
where $\Box$ is the  covariant d'Alembertian.\footnote{More precisely, in \cite{Wetterich:1997bz} was considered a term of the form
$R[\Box +\xi R+{\cal O}(R^2)]^{-1}R$ in Euclidean space.}
Since $\iBox R$ is dimensionless, the associated  constant $g^2$ is also dimensionless. The model however did not produce a viable cosmological evolution~\cite{Wetterich:1997bz}. Deser and Woodard \cite{Deser:2007jk,Deser:2013uya}  considered a more general nonlocal model of the form 
\be\label{GammaDW}
\Gamma_{\rm DW}=\frac{\mplr^2}{2}\int d^4x \sqrt{-g}\,\[ R- Rf(\iBox R)\]\, ,
\ee
with $f(X)$ a dimensionless function (see \cite{Woodard:2014iga} for review). This function  is tuned so to obtain the desired background evolution. Requiring that the cosmological evolution closely mimicks that of $\Lambda$CDM leads to a function well-fitted by~\cite{Deffayet:2009ca}
\be\label{f(X)}
f(X)=0.245\[ \tanh\(0.0350(X+16.5)+0.032 (X+16.5)^2+0.003 (X+16.5)^3\)-1\]\, ,
\ee
for $X\equiv \iBox R <0$. To comply with solar system constraints, the proponents of the model set $f(X)=0$ for $X>0$, a crucial point to which we will come back later. At the level of background evolution the DW model is not predictive, given that $f(X)$ can be chosen to match any desired expansion. However, cosmological perturbations over the background determined by \eq{f(X)} have been studied, and it has been found that the model can fit structure formation data~\cite{Nersisyan:2017mgj,Park:2017zls}.
Similar nonlocal models, involving operators such as $\RMN\iBox\Gmn$, have been discussed in~\cite{Barvinsky:2003kg,Barvinsky:2011hd,Barvinsky:2011rk}.

Both the Wetterich and the Deser-Woodard proposals  involve only dimensionless parameters associated to  the nonlocal term. 
Another interesting possibility is that the nonlocal term involves a  mass scale. Such a mass scale could be generated dynamically in quantum gravity within GR itself (or else be induced in the quantum effective action in extensions of GR where there is already an explicit mass or length scale in the fundamental action, such as massive gravity or theories with extra dimensions). Some indication for the appearance of an IR  mass scale in GR at the non-perturbative level comes from lattice gravity~\cite{Hamber:1999nu,Hamber:2004ew,Hamber:2005dw,Hamber:2015jja} and a recent simulation using causal dynamical triangulations~\cite{Knorr:2018kog}. 
Recently it has also been pointed out that  a dynamical IR mass scale  also  appears using exact renormalization group equations, as a reflection of the instability of the conformal mode in Euclidean space~\cite{Morris:2018mhd}.  A further hint comes from $N=1$ supergravity compactified on a circle, which is a model that allows one to perform non-perturbative computations that otherwise,  in general, are not feasible. In this setting, an IR mass scale emerges because of instanton effects, from the running of the coefficient of the Gauss-Bonnet term \cite{Tong:2014era}.
While these results must  be taken  with a grain of salt, due to the difficulty of dealing with non-perturbative effects in gravity, still, they can at least be considered as a proof of principle of the fact that a mass scale could be dynamically generated in quantum gravity. 

At a purely phenomenological level, a first nonlocal model associated to a mass scale was proposed by Arkani-Hamed,  Dimopoulos,  Dvali, and Gabadadze~\cite{ArkaniHamed:2002fu} to introduce the 
degravitation idea,  and consisted in modifying the Einstein equations  into
\be\label{degrav}
\(1-\frac{m^2}{\Box}\)\Gmn=8\pi G\,\Tmn\, ,
\ee
where $m$ is the new mass scale.\footnote{Actually, to preserve causality, this must be understood as the equation of motion derived from a quantum effective action for the in-in vacuum expectation value of the metric, which automatically ensures that the Green's function in the $\iBox$ operator is the retarded one; see \cite{Maggiore:2016gpx,Belgacem:2017cqo}  for discussion. The same will hold for similar equations, such as \eq{RT} that defines the RT model.}
However, a  drawback of \eq{degrav} is that the energy-momentum tensor is not automatically conserved, since in curved space   $[\n^{\mu},\iBox]\neq 0$ and therefore  the Bianchi identity $\n^{\mu}\Gmn=0$ no longer ensures $\n^{\mu}\Tmn=0$. In \cite{Jaccard:2013gla} it was however observed that it is possible to cure this problem by making use of the fact that 
any symmetric tensor $\Smn$ can be decomposed as 
\be\label{splitSmn}
S_{\mu\nu}=S_{\mu\nu}^{\rm T}+\frac{1}{2}(\n_{\mu}S_{\nu}+\n_{\nu}S_{\mu})\, , 
\ee
where $S_{\mu\nu}^{\rm T}$ is the transverse part of $\Smn$, 
$\n^{\mu}S_{\mu\nu}^{\rm T}=0$. Such a  decomposition can be performed in a generic curved space-time~\cite{Deser:1967zzb,York:1974}. One can then modify \eq{degrav}  into
\be\label{GmnT}
\Gmn -m^2\(\iBox\Gmn\)^{\rm T}=8\pi G\,\Tmn\, ,
\ee
where the superscript ``T" denotes the extraction of the transverse part from of the tensor $\(\iBox\Gmn\)$. Energy--momentum conservation then becomes automatic. However, the  cosmological evolution of this model  turned out to be unstable, already at the background level~\cite{Maggiore:2013mea,Foffa:2013vma}, so this model  is not phenomenologically viable.\footnote{This instability is due to some growing modes in the homogeneous equation for the auxiliary fields used to put the model in a local form; see   \cite{Maggiore:2016gpx,Belgacem:2017cqo} and section~\ref{sect:cosmoRR} for review.}
Then, in \cite{Maggiore:2013mea} a model was  considered  based on the nonlocal equation of motion
\be\label{RT}
\Gmn -\frac{m^2}{3}\(\gmn\iBox R\)^{\rm T}=8\pi G\,\Tmn\, ,
\ee
where  $\iBox$ is rather applied  the Ricci scalar $R$
(the factor $1/3$ is a useful normalization for the mass scale $m$).
We refer to  it as
the ``RT" model, where R stands for the Ricci scalar and T for the extraction of the transverse part.
A closed form for the quantum effective action from which \eq{RT} could be derived is currently not known. This model is however   related to another nonlocal model,  proposed in \cite{Maggiore:2014sia}, and defined by the quantum effective action
\be\label{RR}
\Gamma_{\rm RR}=\frac{\mplr^2}{2}\int d^{4}x \sqrt{-g}\, 
\[R-\frac{m^2}{6} R\frac{1}{\Box^2} R\]\, ,
\ee
that we denote as   the RR model, after the two occurrences of the Ricci scalar in the nonlocal term.
The RT and RR models are related by the fact that they coincide when   linearized over Minkowski space. However beyond linear order, or when linearized over a different background such as FRW, they are different. 

It turns out that, in both the RR and RT models, the auxiliary fields used to put the model in local form  have a stable cosmological evolution at the background level [the condition that ruled out the model (\ref{GmnT})]. In both models the nonlocal term acts   effectively as dark energy, producing accelerated expansion in the recent epoch, and  their cosmological perturbations are well-behaved  in all phases of the cosmic evolution~\cite{Dirian:2014ara}  (e.g. an early inflationary phase, radiation dominance, matter dominance and the recent dark-energy dominated epoch).\footnote{In a paper of our group \cite{Cusin:2016mrr} we incorrectly stated that the RT model is not viable if its evolution is started from an early inflationary phase, because of an instability at the level of scalar perturbations. This statement has been corrected in the v3 arXiv version of that paper, where it is shown that this apparent instability is numerically irrelevant because of the smallness of the scale associated to the nonlocal term compared to the inflationary scale.
The RT model is then perfectly viable even if its evolution is started during an early inflationary phase.} Detailed comparison with CMB, BAO, SNe and structure formation data~\cite{Dirian:2014bma,Dirian:2016puz,Dirian:2017pwp,Belgacem:2017cqo} show that both models fit the data at a level statistically indistinguishable from $\Lambda$CDM. Furthermore, in these models the tensor perturbations propagate at the speed of light, as they should to comply with the limits from the neutron star binary coalescence GW170817, but still with a different ``friction term", which leads to modified GW propagations and distinct predictions that could be tested with LISA or with the Einstein Telescope~\cite{Belgacem:2017ihm,Belgacem:2018lbp}. Thus, these models have a very interesting phenomenology.

As for their physical meaning, it is interesting to observe that  nonlocal terms of this type allow us to write down  mass terms for gauge fields or for some components of the gravitational field,  while preserving gauge or diffeomorphism invariance, respectively. This was  observed in \cite{Dvali:2006su}
for massive electrodynamics, where it was shown that the Proca action for a massive photon,
\be\label{1Lemmass}
S_{\rm Proca}=\int d^4x\[ -\frac{1}{4}F_{\mu\nu}F^{\mu\nu}-\frac{1}{2}m_{\g}^2\, A_{\mu}A^{\mu}\] \, ,
\ee
in which the mass term breaks explicitly the $U(1)$ gauge invariance, is equivalent to the action
\be\label{Lnonloc}
S_{\rm nonlocal}=-\frac{1}{4}\int d^4x\,\[ \Fmn  \FMN-m_{\g}^2\Fmn\frac{1} {\Box}\FMN\]\, ,
\ee
which is nonlocal but gauge-invariant. 
Indeed, in ~\cite{Boucaud:2001st,Capri:2005dy,Dudal:2008sp} it has been found that, in
pure Yang-Mills theory, non-perturbative results obtained from  lattice QCD for the  gluon propagator and the running of the strong coupling constant can be reproduced, at the level of the quantum effective action, by the introduction  of the non-abelian generalization of the above nonlocal term, 
\be\label{Fmn2}
\frac{m^2}{2} \, \int d^4x\,  {\rm Tr}\,\[ F_{\mu\nu} \frac{1}{D^2}F^{\mu\nu}\]\, ,
\ee
(where  $D_{\mu}^{ab}=\d^{ab}\pam-gf^{abc}A_{\mu}^c$ is the covariant derivative).  In this case the nonlocal  term is generated dynamically in the IR  by the strong interaction, and has the meaning of an effective gluon mass. 

The nonlocal terms in \eqs{RT}{RR} have a similar interpretation as a gauge-invariant mass for the conformal mode of the metric~\cite{Maggiore:2015rma,Maggiore:2016fbn}.
Indeed, if we restrict for simplicity the dynamics to the conformal mode $\sigma(x)$, writing 
$\gmn(x)=e^{2\sigma(x)}\emn(x)$, for the  Ricci scalar  we get
$R=-6 e^{-2\sigma}\( \Box\sigma +\pam\sigma\paM\sigma\)$.
Therefore, to linear order in $\sigma$,
$R=-6\Box\sigma +{\cal O}(\sigma^2)$ and, upon integration by parts,
\be\label{m2s2}
 R\frac{1}{\Box^2} R=36 \sigma^2 +{\cal O}(\sigma^3)\, .
\ee
Thus, the  structure $R\Box^{-2}R$ in \eq{RR} 
corresponds to  a nonlocal but diffeomorphism-invariant mass term for the conformal mode, plus  higher-order interaction terms (that are nonlocal even when written in terms of  $\sigma$), that are required to reconstruct a diffeomorphism-invariant quantity.\footnote{Observe that, in contrast, since to linear order $R\iBox R\simeq 36\sigma\Box\sigma$,
the nonlocal term in \eq{GammaWett} corresponds to a kinetic term for the conformal mode.} The same conclusion holds for the RT model since it coincides with the RR model when linearized over Minkowski space (although the higher-order interaction terms  differ). Therefore, as long as we consider the RT model linearized over Minkowski, we can consider \eq{RT} as derived from the RR quantum effective action plus higher-order non-linear terms.
Since  nonlocal terms such as $R\Box^{-2}R$ cannot appear in the fundamental action, but only at the level of the quantum effective action, they describe mass terms generated dynamically by quantum effects. It is quite interesting to observe that  numerical results from causal dynamical triangulations recently presented in \cite{Knorr:2018kog} provide some evidence in favor of a mass term for the conformal mode, of the type described by eqs.~(\ref{RT}) or (\ref{RR}).

Other variants of the RR model have been explored, such as terms in the effective action of the form $m^2\RMN\Box^{-2}\Rmn$ or $m^2C_{\mu\nu\rho\sigma}\Box^{-2}C^{\mu\nu\rho\sigma}$, where
$C_{\mu\nu\rho\sigma}$ is the Weyl tensor~\cite{Cusin:2015rex}. However, the former does not have a stable background cosmological evolution, just as the model (\ref{GmnT}); the latter  does not affect the background evolution in FRW, and has well-behaved scalar perturbations, but it turns out to be unstable at the level of tensor perturbations.  Observe that these structures also give a mass to the helicity-2 mode, so a physical mechanism that  generates only a mass for the conformal mode would not produce them.\footnote{In this context, it is useful to clear up a  misconception that occasionally appears in discussions. The quantum effective action should not be confused with  a low-energy  effective action. The quantum effective action is an object that includes the effects of quantum fluctuations (in principle exactly, if one were able to compute it), and is valid at all energy scales at which the original fundamental action is valid. To compute the quantum effective action one is not integrating out some fields from the theory, but rather one is integrating out their quantum fluctuations, remaining with a functional of the vacuum expectation values of all fields (see 
\cite{Maggiore:2016gpx} for a pedagogical discussion).
While a low-energy effective action in general has all terms consistent with the symmetries of the fundamental action, the quantum effective action will have a very specific structure. For instance, some mode could become massive while others could remain massless. Thus, it makes perfect sense to study a specific nonlocal structure; assuming that non-perturbative effects generate a term proportional to $R\Box^{-2}R$  does not imply that then also a term such as, e.g.,  $\RMN\Box^{-2}\Rmn$ will  automatically appear, even more considering that these terms have different physical meaning.} Given that models that have the physical meaning of a mass for the conformal mode appear to be the viable ones, in \cite{Cusin:2016nzi}  a model with a term $m^2R\Delta_4^{-1}R$ was also studied, where $\Delta_4=\Box^2+2\RMN\n_{\mu}\n_{\nu}-\frac{2}{3}R\Box+\frac{1}{3}\gMN\n_{\mu} R\n_{\nu}$ is called the Paneitz operator. This operator enters the quantum effective action obtained by integrating the conformal anomaly in four dimensions~\cite{Riegert:1984kt,Deser:1999zv}. In flat space it reduces to $\Box^2$, so again this nonlocal term has the meaning of a mass for the conformal mode. The model turns out to have a viable evolution at the level of background, as well as in the scalar perturbations sector. However, the study of its tensor perturbations revealed that  GWs do not propagate at the speed of light, and therefore the $\Delta_4$ model is ruled out by GW170817~\cite{Belgacem:2017cqo}. A review of the conceptual aspects and phenomenological consequences of these nonlocal models associated to a mass scale is given in   \cite{Maggiore:2016gpx,Belgacem:2017cqo} (see also \cite{Foffa:2013sma,Nesseris:2014mea,Conroy:2014eja,Barreira:2015fpa,Barreira:2015vra,Nersisyan:2016hjh,Vardanyan:2017kal,Amendola:2017qge}  for further works).

These considerations show that the construction of a phenomenologically viable nonlocal model is a highly non-trivial task. In this paper, elaborating on the discussion in~\cite{Barreira:2014kra}, we examine another constraint on these models, coming from the bound on the time variation of Newton's constant obtained from Lunar Laser Ranging (LLR). We will find that, within some caveats to be discussed below, this bound rules out both the DW model and the RR model, while the RT model survives.

The paper is organized as follows. In section~\ref{sect:limitLLR} we review the current status of the limit on $\dot{G}/G$ from LLR, and we compare with a first `naive' analysis for the RR model (while we will see that the RT model is not concerned by the bound, and passes  the LRR test automatically). In section~\ref{sect:match} we put the  analysis for the RR model on firmer grounds, by examining different ways of matching a cosmological solution at large scales to a static solution at short scales, and we will   conclude that the RR model is indeed ruled out. In section~\ref{sect:noscreeningDW} we analyze a screening mechanism that was proposed for the Deser-Woodard model. We show that this mechanism actually does not work and that the  Deser-Woodard model is also ruled out. We present our conclusions in section~\ref{sect:conclusions}. Some technical material is relegated in three appendices.
We use units $c=1$ and MTW conventions for the metric signature, Riemann tensor, etc.

\section{Limits from $\dot{G}/G$ and modified gravity}\label{sect:limitLLR}

Lunar Laser Ranging provides a stringent limit on the time variation of Newton's constant. The most recent result is~\cite{Hofmann:2018myc}
\be\label{dotGsuG}
\frac{\dot{G}}{G}=(7.1\pm 7.6)\times 10^{-14}\, {\rm yr}^{-1}\, .
\ee
Note that the LLR limit improves significantly with observation time and the above limit is one order of magnitude  stronger than the previously published limit $(4\pm 9) \times 10^{-13}\, {\rm yr}^{-1}$ \cite{Williams:2004qba}, which was also used in \cite{Barreira:2014kra} to discuss constraints on nonlocal gravity. The cosmological implications of this result become more evident if we re-express this value in terms of the Hubble parameter today $H_0=h_0\times 100 \, {\rm km}\, {\rm s}^{-1}\, {\rm Mpc}^{-1}\simeq 
h_0\times (9.777 752\, {\rm Gyr})^{-1}$. Then, \eq{dotGsuG} reads
\be\label{dotGsuGH0}
\frac{\dot{G}}{G}= (0.99\pm 1.06)\times 10^{-3}\, \( \frac{0.7}{h_0}\) \, H_0\, .
\ee
Quite generally, in modified gravity models Newton's constant becomes time dependent on cosmological scale.  The scale for the time variation today is  given by $H_0$, so, generically, on cosmological scales one finds
$\dot{G}/G\simeq H_0$, with a coefficient typically of order one. If, in a given modified gravity model, such a result persists down to  the scale of the solar system and of the Earth-Moon system, then the  model is ruled out by
\eq{dotGsuGH0}. A modified gravity models can however be immune to this bound, either because  it simply does not induce a time-dependence in Newton's constant at small scales (we will see that this is the case for the RT model),  or because it exhibits a non-linear screening mechanism, such as the Vainshtein mechanism or the chameleon mechanism (see e.g.~\cite{Joyce:2014kja} for review). 

\subsection{Effective Newton's constant in modified gravity}\label{sect:defGeff}

Before specializing to the non-local models that will be the main focus of our work, it is useful to discuss in more general terms why in modified gravity Newton's constant is typically modified, and what is the relation between the effective Newton  constant measured on cosmological scales and that measured from the motion of test masses on shorter scales, as in the case when we take the Moon as a test mass in the field of the Earth.
Generically, in modified gravity the Einstein equations become
\be\label{modifEin1}
\Gmn+\Delta\Gmn=8\pi G\Tmn\, ,
\ee
where $\Delta\Gmn$ can depend on extra fields of the theory, e.g. on some scalar fields in the case of scalar-tensor theories. As we will recall below, in non-local gravity $\Delta\Gmn$  rather depends on non-local functional of the metric, and can be rewritten as a local expression in terms of some auxiliary fields. The set of equations of the theory is then completed by the equations of motion of the extra fields (or, in non-local gravity, by equations that follow from the definition of the auxiliary fields, see below). Suppose that, among the various terms in  $\Delta\Gmn$, there will be terms proportional to the Einstein tensor $\Gmn$ itself, so 
$\Delta\Gmn=\a(\phi)\Gmn+{\cal G}_{\mu\nu}$, where $\a(\phi)$ is some function of the various fields or auxiliary fields of the theory, here denoted collectively as $\phi$ (and of the metric, that we do not write explicitly), and ${\cal G}_{\mu\nu}$ denotes the remainder, that is not proportional to $\Gmn$. Of course, this assumes a choice of basis in the space of operators; in particular, it is convenient to  take $\Gmn$ and $\gmn$ as independent (rather than, say, $\Gmn$ and $\Rmn$) because the former, as we will see, contributes to the effective Newton  constant, while terms proportional to $\gmn$ contributes to a cosmological constant or a potential term, so this separation has a more clear physical interpretation.
\Eq{modifEin1} can then be rewritten as
\be\label{modifEin2}
\Gmn=8\pi G_{\rm eff} \(\Tmn-\frac{1}{8\pi G}{\cal G}_{\mu\nu}\)\, ,
\ee
where 
\be\label{modifEinGeff}
G_{\rm eff}(\phi)=\frac{G}{1+\a(\phi)}
\ee
plays the role of an effective Newton's constant. Whether this separation of $\Delta\Gmn$ into   $\a(\phi)\Gmn$ and ${\cal G}_{\mu\nu}$ is useful or not depends on the context. For instance, at the level of background evolution, \eq{modifEin2}  produces a modified Friedmann equation of the form 
\be\label{modifEin3}
H^2=\frac{8\pi}{3} G_{\rm eff}(\phi) \[ \rho+\tilde{\rho}(\phi)\]\, ,
\ee
where $\tilde{\rho}(\phi)$ is the contribution from ${\cal G}_{00}$. However, the time dependence of 
$\tilde{\rho}(\phi)$ will in general be as important as that in $G_{\rm eff}(\phi)$, so this separation is not very useful. It is more meaningful to put all the contribution from  $\Delta\Gmn$ together, defining an effective dark energy density $\rho_{\rm DE}(\phi)$ from $G_{\rm eff}(\phi) \[ \rho+\tilde{\rho}(\phi)\]=
G\[ \rho+\rho_{\rm DE}(\phi)\]$, so that \eq{modifEin3} reads
\be\label{modifEin4}
H^2=\frac{8\pi G}{3} \[ \rho+\rho_{\rm DE}(\phi)\]\, .
\ee
At the level of cosmological perturbations, this separation can however be more useful. Indeed, working with a perturbed FRW metric 
\be\label{Poissongauge0}
ds^2=-(1+2\Psi) dt^2+(1+2\Phi)a^2(t)d\vx^2\, ,
\ee
where  $\Phi(t,\vx)$ and $\Psi(t,\vx)$ are the  usual  Bardeen potentials,
the linearization of the $(00)$ component of \eq{modifEin2}, for modes  inside the horizon,  gives a modified Poisson equation of the form
\be\label{modifEin5}
\n^2\Phi=-4\pi G_{\rm eff}(\phi)  a^2\, \[\delta\rho+\delta\tilde{\rho}(\phi)\]\, ,
\ee
where again $\delta\tilde{\rho}(\phi)$ is the contribution from $\delta{\cal G}_{00}$. However, in several modified gravity models it happens that the term $\delta\tilde{\rho}(\phi)$ is negligible for modes well inside the horizon. As we will recall below, this is in particular the case for the RR and RT   nonlocal models. In that case, $G_{\rm eff}$ defined from \eq{modifEin2} captures the main effect on the growth of the perturbations [in general, together with a second indicator such as the gravitational slip $(\Phi+\Psi)/\Phi$].

However, the situation where this separation is most useful is when we study a modified gravity model at the scale of the solar system or the Earth-Moon system, as for LLR. Indeed, in this case, the term
${\cal G}_{\mu\nu}/(8\pi G)$ in \eq{modifEin2} is totally negligible with respect to $\Tmn$. For instance, its $(00)$ component is of the order of the energy density of dark energy and, at the solar system scales, as an additive contribution its effect on the dynamics is totally irrelevant compared to $T_{00}$, which is the energy density $\rho$ of the bodies, such as the Earth, that dominate the dynamics at the scales relevant for LLR. In contrast, the effective Newton  constant gives a multiplicative contribution to $\rho$. Even if numerically tiny, the time dependence that it induces can be tested through the very accurate LLR bound (\ref{dotGsuGH0}).

To sum up, in modified gravity the effective Newton constant relevant for comparison with LLR can be read directly from the modified Einstein equations,  by selecting the terms proportional to $\Gmn$ in $\Delta\Gmn$, and its functional dependence on the fields $\phi$, given by \eq{modifEinGeff}, is   the same as that of the effective Newton constant that governs linear cosmological perturbation theory through the modified Poisson equation (\ref{modifEin5}). In general, different screening mechanisms will be possible depending on the specific dynamics of the fields $\phi$, which could suppress the value of $\a(\phi)$ in the regime relevant for solar system tests.
We next examine some concrete cases, namely the RT, RR and DW models.

\subsection{RT model}

Let us  first recall how the RT model can be recast in local form~\cite{Maggiore:2013mea}. We begin   by introducing an auxiliary field $U=-\iBox R$ and we define  
$S_{\mu\nu}=-U\gmn=\gmn \iBox R$.\footnote{The notation that has been traditionally used for the RR and RT model is different from the one that is standard in the DW model. In particular, the auxiliary field called $U$ in the RT and RR models correspond to $-X$ in the notation used for the DW model.}
We then split this tensor into a  transverse part $S_{\mu\nu}^{\rm T}$, which by definition satisfies $\n^{\mu}S_{\mu\nu}^{\rm T}=0$, and a longitudinal part, 
\be\label{splitSmnRT}
S_{\mu\nu}=S_{\mu\nu}^{\rm T}+\frac{1}{2}(\n_{\mu}S_{\nu}+\n_{\nu}S_{\mu})\, . 
\ee
Thus, the nonlocal quantity $S_{\mu\nu}^{\rm T}=(\gmn \iBox R)^{\rm T}$ that appears in \eq{RT} can be written as
\be
(\gmn \iBox R)^{\rm T}=-U\gmn-\frac{1}{2}(\n_{\mu}S_{\nu}+\n_{\nu}S_{\mu})\, ,
\ee
in terms of an auxiliary scalar field $U$ and an auxiliary four-vector field $S_{\mu}$. Then the equation of motion (\ref{RT}) can be rewritten as the coupled system of equations
\bees
\Gmn +\frac{m^2}{6}\, \(2U\gmn + \n_{\mu}S_{\nu}+\n_{\nu}S_{\mu}\)&=&8\pi G\,\Tmn\, ,
\label{v2loc1}\\
\Box U&=&-R\, ,\\
(\d^{\mu}_{\nu}\Box +\n^{\mu}\n_{\nu})S_{\mu}&=&-2\pan U\, , \label{panU}
\ees
where the last equation is obtained by taking the divergence of \eq{splitSmn}. 
We now observe that \eq{v2loc1} has the form (\ref{modifEin1}), but with a $\Delta\Gmn$ in which there 
no term proportional to $\Gmn$. Therefore, according to the discussion in section~\ref{sect:defGeff}, the effective Newton constant relevant for LLR is equal to $G$.

This can be confirmed by expanding the equations to linear order over FRW, and reading the effective Newton constant from the Poisson equation.
For the RT model this computation  has been performed in \cite{Nesseris:2014mea,Dirian:2014ara}. The result is that, for sub-horizon modes,
\be\label{GeffGRTlargek}
\frac{G_{\rm eff}}{G}=1+{\cal O}\(\frac{1}{\hat{k}^2}\)\, ,
\ee
where $\hat{k}=k/(aH)$ is the physical momentum $k/a$ divided by $H$. This confirms again that,
for sub-horizon modes, 
$G_{\rm eff}$ reduces to $G$ and the correction, for the modes relevant to  LLR, is totally negligible. 
Thus the RT model passes the test (\ref{dotGsuGH0}).

\subsection{RR model}\label{sect:LRRvsRR}

We next consider the RR model. Let us recall that, for this model, it is convenient to introduce two auxiliary fields $U$ and $S$ defined from 
$U=-\iBox R$ and $S=-\iBox U$~\cite{Maggiore:2014sia}. Then, the equation of motion of the theory become equivalent to the following coupled system
\bees
\Gmn&=&\frac{m^2}{6} K_{\mu\nu}+8\pi G\Tmn\, ,\label{Gmn}\\
\Box U&=&-R\,, \label{BoxURR}\\
\Box S &=&-U\, ,\label{BoxS}
\ees
where  
\be\label{defKmn}
K_{\mu\nu}=2S\Gmn-2\n_{\mu}\pan S -2U\gmn+\gmn\parho S\paR U -\frac{1}{2}\gmn U^2-(\pam S\pan U+\pan S\pam U)\, .
\ee
Now in the $K_{\mu\nu}$ there is a term proportional to $\Gmn$. As a consequence, the effective Newton constant relevant for LLR is given by
\be\label{Geff1LLR}
G_{\rm eff} =  \frac{G}{ 1-(1/3)m^2 S}\,.
\ee
Again, we can compare the the result obtained  by
expanding \eqst{Gmn}{BoxS} to linear order over FRW. The result, for sub-horizon modes, is~\cite{Dirian:2014ara,Barreira:2014kra}
\be\label{Geff1suk2}
\frac{G_{\rm eff}(t)}{G} =  \[1-\frac{1}{3}m^2 \bar{S}(t)\]^{-1}\, \[ 1 +{\cal O}\(\frac{1}{\hat{k}^2}\)\]\, ,
\ee
where $\bar{S}(t)$ is the background cosmological solution for the auxiliary field $S$. 
\Eq{Geff1suk2} shows that the effective Newton constant that enters the Poisson equation  for the cosmological perturbations  has a dependence on time through the background value of the auxiliary field $\bar{S}(t)$. The effective Newton constant relevant for LLR, given by \eq{Geff1LLR}, has the same functional dependence on $S$, as expected from the discussion in section~\ref{sect:defGeff},
so in this case there is a potential problem with LLR, as  observed in
\cite{Barreira:2014kra}. Of course, the crucial issue, also recognized in  \cite{Barreira:2014kra}, is that  it is not  evident whether the cosmological time-dependent solution for $S$ is still appropriate at solar system or LLR scales, where one could rather imagine that the cosmological solution matches to a static solution $S_{\rm static}(r)$, i.e. it is not obvious whether, when we apply \eq{Geff1LLR} to LLR, we have the right to replace $S$ by the cosmological solution $\bar{S}(t)$.

We will examine this issue in detail in the following sections. For the moment, if one just simply assumes that \eq{Geff1suk2} is applicable at LLR scales and uses the background solution for the $S$ field obtained from the numerical integration of the equations of motion
(\ref{Gmn})--(\ref{BoxS}), one finds $\dot{G}/G\simeq 0.12\, H_0$. As observed in \cite{Barreira:2014kra}, this already exceed by a factor ${\cal O}(10)$ the bound $(4\pm 9) \times 10^{-13}\, {\rm yr}^{-1}$ \cite{Williams:2004qba}, that in terms of $H_0$ reads $\dot{G}/G\simeq (5\pm 12)\times 10^{-3} H_0 (0.7/h_0)$. The  more recent bound (\ref{dotGsuGH0}) is then exceeded by a factor ${\cal O}(100)$.

Actually, as we will recall in section~\ref{sect:cosmoRR}, the evolution in the RR model is also characterized by one free parameter, which is the initial value $u_0$ of the auxiliary field $U$, at an initial time set deep in radiation dominance, and represents a marginal direction in parameter space (while, in the space of initial conditions $\{U(t_i),\dot{U}(t_i),S(t_i),\dot{S}(t_i)\}$, the  remaining three directions correspond to irrelevant directions in parameter space). The above result for $\dot{G}/G$ was obtained
for the ``minimal model", defined by $u_0=0$. Larger values of $u_0$ bring the evolution of the model closer and closer to that of $\Lambda$CDM, and could be naturally generated by a phase of primordial inflation. Indeed, $U$ grows during inflation in such a way that, starting from a value of order one at the beginning of inflation, at the end of inflation it reaches a value $U\simeq 4\Delta N$, where $\Delta N$ is the number of efolds. This value   can then be taken as the initial value  $u_0$ for the subsequent evolution during radiation dominance. For $\Delta N=60$ this gives  a large value $u_0\simeq 240$. It is therefore interesting to see if the bound (\ref{dotGsuGH0}) can be avoided with large values of $u_0$. We have therefore computed $\dot{G}/G$ from the numerical evolution for several values of $u_0$. Using our standard notations (see  section~\ref{sect:cosmoRR})
$\gamma\equiv m^2/(9H_0^2)$ and $V\equiv H_0^2S$, parametrizing the time evolution in terms of $x\equiv \ln a(t)$, and denoting by a prime the derivative with respect to $x$,  for modes well inside the horizon we have
\be
\frac{G_{\rm eff}(t)}{G} =   \frac{1}{1-3\gamma\bar{V}(t)}\, ,
\ee
and therefore
\be
\frac{\dot{G}_{\rm eff}}{G_{\rm eff}}=H(t)\, \frac{3\gamma\bar{V}'}{1-3\gamma\bar{V}}\, .
\ee
Thus, at the present time, $\(\dot{G}_{\rm eff}/{G_{\rm eff}}\)_0=\alpha H_0$, where

\begin{figure}[t]
\includegraphics[width=0.45\columnwidth]{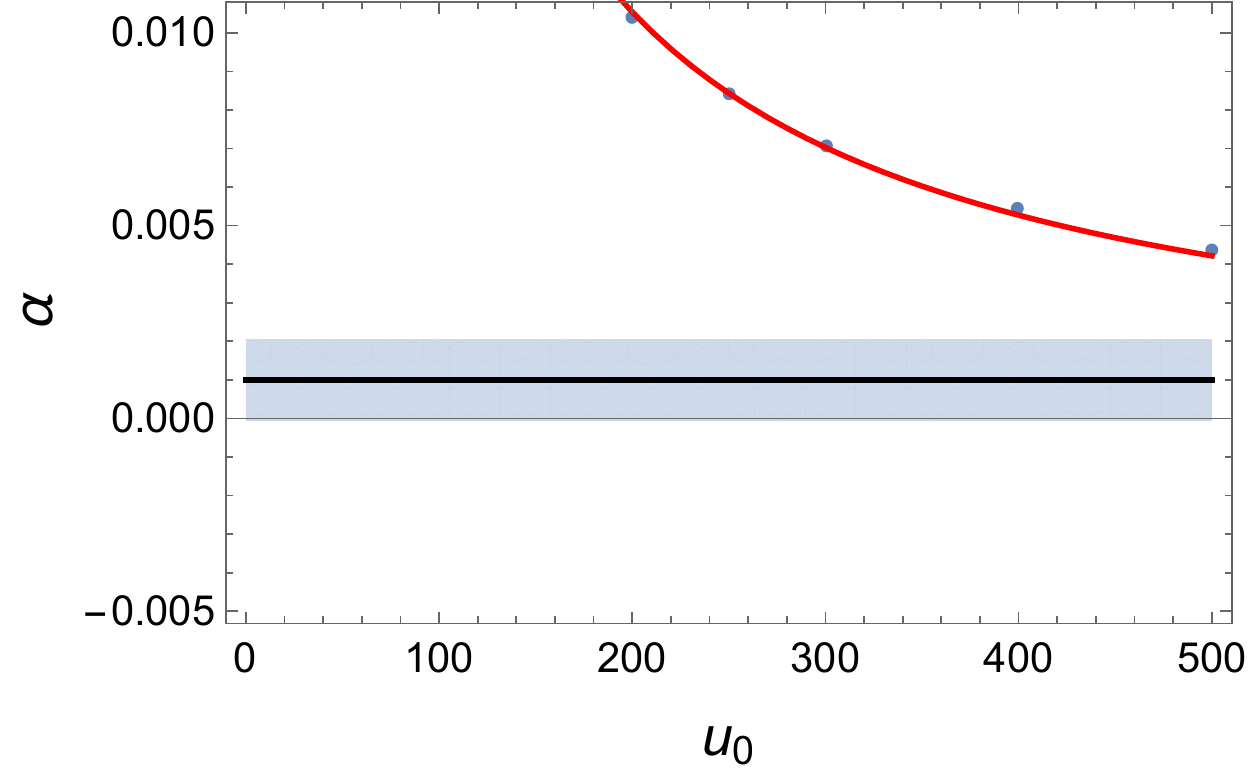}
\caption{\label{fig:alpha_vs_u0} The value of $\alpha$ computed numerically for several values of $u_0$ (blue dots) compared with a fit  of the form $\alpha \simeq 2.11/u_0$ (red line) and the observational limit  (\ref{dotGsuGH0}) (mean value as black solid line, $1\sigma$ region shaded).}
\end{figure}

\be
\alpha = \(\frac{3\gamma\bar{V}'}{1-3\gamma\bar{V}}\)_0\, ,
\ee
and the subscript $0$ denotes the present time. This quantity is easily computed from the solution for $\bar{V}(x)$ obtained from the numerical integration of the equations of motion of the model, specialized to a FRW background \cite{Maggiore:2014sia,Maggiore:2016gpx,Belgacem:2017cqo}, that we will recall in section~\ref{sect:cosmoRR}.
In Fig.~\ref{fig:alpha_vs_u0} we compare the numerical results for $\alpha$, obtained for different values of the initial condition $u_0$,  with the observational limit  (\ref{dotGsuGH0}). We see that, under the assumption that \eq{Geff1suk2} is indeed still valid at the scale of the Earth-Moon system, the RR model is ruled out
even for very large values of $u_0$, say $u_0\simeq 500$. Of course, with a sufficiently large value of $u_0$,  the model would eventually comply with the limit. According to the fit $\alpha \simeq 2.11/u_0$ (red line), this requires $u_0\, \gsim\, 2000$. However, such values becomes unnaturally large even from the point of view of a previous inflationary phase. Furthermore, the model would become so close to $\Lambda$CDM that it would become basically indistinguishable from $\Lambda$CDM in its cosmological predictions.

The question is therefore whether the extrapolation of \eq{Geff1suk2} from cosmological scales down to the scale of the Earth-Moon system is justified. We address this issue in  section~\ref{sect:match}.

\subsection{DW model}

We next consider  the DW model. The variation of \eq{GammaDW} gives a modified Einstein equation of the form (\ref{modifEin1}), with~\cite{Deser:2007jk,Deser:2013uya}
\bees
\Delta\Gmn&=&\[ \Gmn +\gmn\Box-\n_{\mu}\n_{\nu}\]\left\{
f(\iBox R)+\iBox\[ Rf'(\iBox R) \] \right\} \nn\\
&&+\[ \delta_{\mu}^{(\rho}\delta_{\nu}^{\sigma )}-\frac{1}{2}\gmn\gRS\]
\parho f(\iBox R)\pas \(\iBox\[ Rf'(\iBox R) \] \)\, ,
\ees
where $f'(X)=\pa f/\pa X$. The model can be written in local form by introducing two auxiliary fields $X=\iBox R$ and $Y=\iBox\[ Rf'(X)\]$, so that the equations for the auxiliary fields are $\Box X=R$ and $\Box Y=R f'(X)$.\footnote{In the literature on the DW model the auxiliary field $Y$ is denoted by $U$. We change the notation to avoid confusion with the auxiliary field denoted by $U$ in the RR and RT models, that rather corresponds to $-X$.} Picking the term proportional to $\Gmn$ one finds the effective Newton constant~\cite{Woodard:2014iga}
\bees
G_{\rm eff}&=&\frac{G}{1+f(\iBox R)+\iBox\[ Rf'(\iBox R) \] }\nn\\
&=&\frac{G}{1+f(X)+Y}\, .\label{GeffDW}
\ees
The same result can be found by writing the Poisson equation for the perturbations around FRW in the large-$k$ 
limit, which gives~\cite{Nersisyan:2017mgj,Park:2017zls}
\be
k^2\[1+f(\bar{X})+\bar{Y}\]\Phi_k+ \frac{1}{2} k^2 \[\delta U_k+f'(\bar{X})\delta X_k\]=4\pi G a^2\delta_k\, ,
\ee
where as usual the overbar denotes the field evaluated on the background cosmological solutions.
Note that in this case, contrary to the RR and RT models, besides an effective Newton constant there is also an extra contribution  proportional to $[\delta U+f'(\bar{X})\delta X]$, that does not vanish in the sub-horizon limit, i.e. the term $\delta\tilde{\rho}(\phi)$ in \eq {modifEin5} is not suppressed for large $k$.

The time dependence of the auxiliary fields $X$ and $Y$ therefore potentially induces a time dependence in the Newton constant that govern the dynamics at LLR scales. 
In the attempt at introducing a screening mechanism, the proponents of the model set $f(X)=0$ when $X>0$, a point that will be discussed in detail in section~\ref{sect:noscreeningDW}. 

\section{Matching of cosmological and static solutions}\label{sect:match}

A first natural reaction to the  result found in section~\ref{sect:LRRvsRR} for the RR model is that, after all, the FRW metric has no relevance for the Earth-Moon system, which is rather governed by a static \Sch metric. However, as correctly observed in 
\cite{Barreira:2014kra}, the issue is more subtle. The complete metric will, of course, be one that, at the largest cosmological scales, reduces to  FRW and, as we go down in scales, is gradually affected by the various structures and inhomogeneities of the Universe (superclusters, clusters, galaxies, etc.), until at the solar system scale is  equal to the \Sch metric generated by the Sun (plus the corrections from  the other relevant bodies), and finally  at the Earth-Moon scale it  reduces to the static metric generated by the Earth. The auxiliary fields $U$ and $S$ of the RR model satisfy \eqs{BoxURR}{BoxS}, where the $\Box$ operator and the Ricci scalar are computed in this metric. At the largest scales, where the metric is given by FRW, we get the corresponding cosmological solutions  $U_{\rm cosmo}(t)$ and $S_{\rm cosmo}(t)$, that depend on time but not on $\vx$, and whose explicit form we will recall in section~\ref{sect:cosmoRR}. At solar system or Earth-Moon scales, we can study the solution of \eqs{BoxURR}{BoxS} in a static \Sch metric. In this background we can look for static solutions, that we denote here by $U_{\rm static}(r)$ and 
$S_{\rm static}(r)$, and that will be discussed in section~\ref{sect:staticRR} and app.~\ref{App:Sch}, extending the discussion 
of refs.~\cite{Maggiore:2014sia,Kehagias:2014sda}. These solutions are static by construction since they have been found by looking for solutions with no time dependence. However, it is not obvious whether, in the full metric that interpolates between the FRW metric at the largest cosmological scales and the \Sch metric at the Earth-Moon scale, the complete solution for  $S$ will correspondingly interpolate between $S_{\rm cosmo}(t)$ and $S_{\rm static}(r)$. It is in principle possible that at the Earth-Moon scale the solution for $S$ could still retain a time dependence,   inherited from the time evolution at large scales (and which is lost by definition when searching for a purely static solution, matched at infinity to  flat space). If this is the case, then Newton's constant will indeed have a  time dependence at the Earth-Moon scale  relevant for the LLR observations. Indeed, this is precisely what happens in models, such as galileons or k-essence, in which a field $\varphi$ has a shift symmetry $\varphi\ra \varphi+{\rm const}$. In this case, thanks to the shift symmetry, near the present epoch $t_0$ the equation of motion of the field admits a solution with separation of variables of the form 
$\varphi(t,r)=\varphi_{\rm static}(r)+\varphi_{\rm cosmo}(t_0)+(t-t_0)\dot{\varphi}_{\rm cosmo}(t_0)$~\cite{Babichev:2011iz}. The RR model does not have such a shift symmetry, so the applicability of this result is not obvious, but, still, this shows that the issue is non-trivial.

An equivalent way of posing the problem is to follow the evolution of the fields $S$ and $U$ starting from radiation dominance (RD), before structure formation. At that time the FRW metric holds everywhere. The results that we will review in section~\ref{sect:cosmoRR} reveal that during RD the auxiliary fields still show no significant time evolution. However, as soon as we enter in matter dominance (MD), the $U$ field begins to
evolve with time. It is not clear what would stop the time evolution of these fields in the regions where the metric is dominated by the contribution of massive bodies, but still the overall gravitational field is weak, which is the case everywhere except in the vicinity of compact objects such as black holes or neutron stars.

To understand the problem, we need to study the equations for $U$ and $S$ in a metric that interpolates between the cosmological and the static solution. 
We begin however by first recalling, in the next two subsections, the form of the cosmological and static solutions for the auxiliary fields in the RR model.

\subsection{Cosmological  solutions for the RR model}\label{sect:cosmoRR}

Let us recall the form of the cosmological solutions for  the RR model~\cite{Maggiore:2014sia,Dirian:2014ara,Belgacem:2017cqo}. We consider a flat FRW metric  
$ds^2=-dt^2+a^2(t)d\vx^2$,
and we look for cosmological background solutions for the auxiliary field that depend only on time, 
$U_{\rm cosmo}(t)$ and $S_{\rm cosmo}(t)$ in the notation of the above discussion. For ease of notation, however, in this subsection we write   $U_{\rm cosmo}(t)=\bar{U}(t)$ and  $S_{\rm cosmo}(t)=\bar{S}(t)$, and in general we use an overbar for quantities  corresponding to the background cosmological evolution.
We define $h(t)=H(t)/H_0$, where  $H(t)=\dot{a}/a$ and $H_0$ is the present value of the Hubble parameter, we use $x=\ln a(t)$ to parametrize the temporal evolution, and we denote $df/dx\equiv f'$. Observe that $U$ is dimensionless while $S=-\iBox U$ has dimension of  $({\rm length })^2$.
To show that the nonlocal term  gives rise to  an effective dark energy density it is convenient 
to introduce the dimensionless field $W(t)=H^2(t)S(t)$. 
The equations of motion
(\ref{Gmn})--(\ref{BoxS}) then become
\bees
&&h^2(x)=\Omega_M e^{-3x}+\Omega_R e^{-4x}+\g \bar{Y}[\bar{U},\bar{V}]\, ,\label{syh}\\
&&\bar{U}''+(3+\zeta) \bar{U}'=6(2+\zeta)\, ,\label{syU}\\
&&\bar{W}''+3(1-\zeta) \bar{W}'-2(\zeta'+3\zeta-\zeta^2)\bar{W}= \bar{U}\, ,\label{syW}
\ees
where $\oma, \ora$ are the present values of $\rho_M/\rho_{\rm tot}$ and $\rho_R/\rho_{\rm tot}$, respectively, 
$\gamma= m^2/(9H_0^2)$, $\zeta=h'/h$ and $\bar{Y}$ is a known functional of $\bar{U},\bar{V}$, whose explicit expression is
$\bar{Y}\equiv (1/2)\bar{W}'(6-\bar{U}') +\bar{W} (3-6\zeta+\zeta \bar{U}')+(1/4)\bar{U}^2$.
In this form, one sees that  there is an effective dark energy density $\rde=\rho_0\gamma \bar{Y}$ where, as usual, 
$\rho_0=3H_0^2/(8\pi G)$. Actually, to perform the numerical integration of these equations, and also to study the cosmological perturbations, it can be more convenient to use a variable $V(t)=H_0^2S(t)$ instead of $W(t)=H^2(t)S(t)$. Then 
\eqst{syh}{syW} become
\bees
&&h^2(x)=\frac{\Omega_M e^{-3x}+\Omega_R e^{-4x}+(\g/4) \bar{U}^2}
{1+\g[-3\bar{V}'-3\bar{V}+(1/2)\bar{V}'\bar{U}']}\,,\label{syh2}\\
&&\bar{U}''+(3+\zeta) \bar{U}'=6(2+\zeta)\, ,\label{syU2}\\
&&\bar{V}''+(3+\zeta) \bar{V}'=h^{-2} \bar{U}\label{syV}\, .
\ees
Given the initial conditions, that we will set for definiteness deep in RD, the numerical integration of these equations is straightforward. The initial conditions on the auxiliary fields $U$ and $V$ (or, equivalently, $U$ and $W$), are in one-to-one correspondence with the solutions of the homogeneous equations associated to
\eqs{syU}{syW}, i.e. 
\bees
&&\bar{U}''+(3+\zeta) \bar{U}'=0\,  ,\\
&&\bar{W}''+3(1-\zeta) \bar{W}'-2(\zeta'+3\zeta-\zeta^2)\bar{W}=0\, .
\ees 
In any given cosmological  epoch $\zeta$ has an approximately constant value $\zeta_0$, with  
$\zeta_0=\{0,-2,-3/2\}$  in de~Sitter (dS),  RD and  MD, respectively. Taking $\zeta$ constant  the homogeneous solutions are obtained analytically,
\bees
\bar{U}_{\rm hom}(x)&=& u_0+u_1 e^{-(3+\zeta_0)x}\, ,\\
\bar{W}_{\rm hom}(x)&=&w_0e^{-(3-\zeta_0)x}+w_1 e^{2\zeta_0x}\, .
\ees
During the whole cosmological evolution  $-2\leq \zeta_0\leq 0$, so  $u_1,w_0$ and $w_1$ parametrize 
 irrelevant directions in  parameter space. Note also that there is no exponentially growing solution, so no instability at this level of the analysis.\footnote{This is the condition that was not fulfilled by the model (\ref{GmnT}).} In contrast, $u_0$ parametrizes a marginal direction in parameter space and can be taken as a free parameter of the model. To have an idea of a natural value for this parameter we observe, from \eq{syU2} with $\zeta_0=0$, that, during a  pre-existing phase of de~Sitter  inflation, $\bar{U}(x)=4x$. Then, if $\bar{U}(x)={\cal O}(1)$ at the beginning of inflation,
by the end of inflation $\bar{U}(x)\simeq 4\Delta N$, where $\Delta N$ is the number of inflationary efolds. Thus $u_0$, which is the initial value for the subsequent evolution starting from RD,  could naturally be expected to be of order $4\Delta N$. For typical inflationary models $\Delta N={\cal O}(50-60)$, so inflation could naturally produce a value of $u_0={\cal O}(200-250)$. We can then study the evolution for the simplest choice $u_0=0$ [which is representative of any value $u_0={\cal O}(1)$] and for a large value of $u_0$, say $u_0=200$. 

\begin{figure}[t]
\centering
\includegraphics[width=0.45\columnwidth]{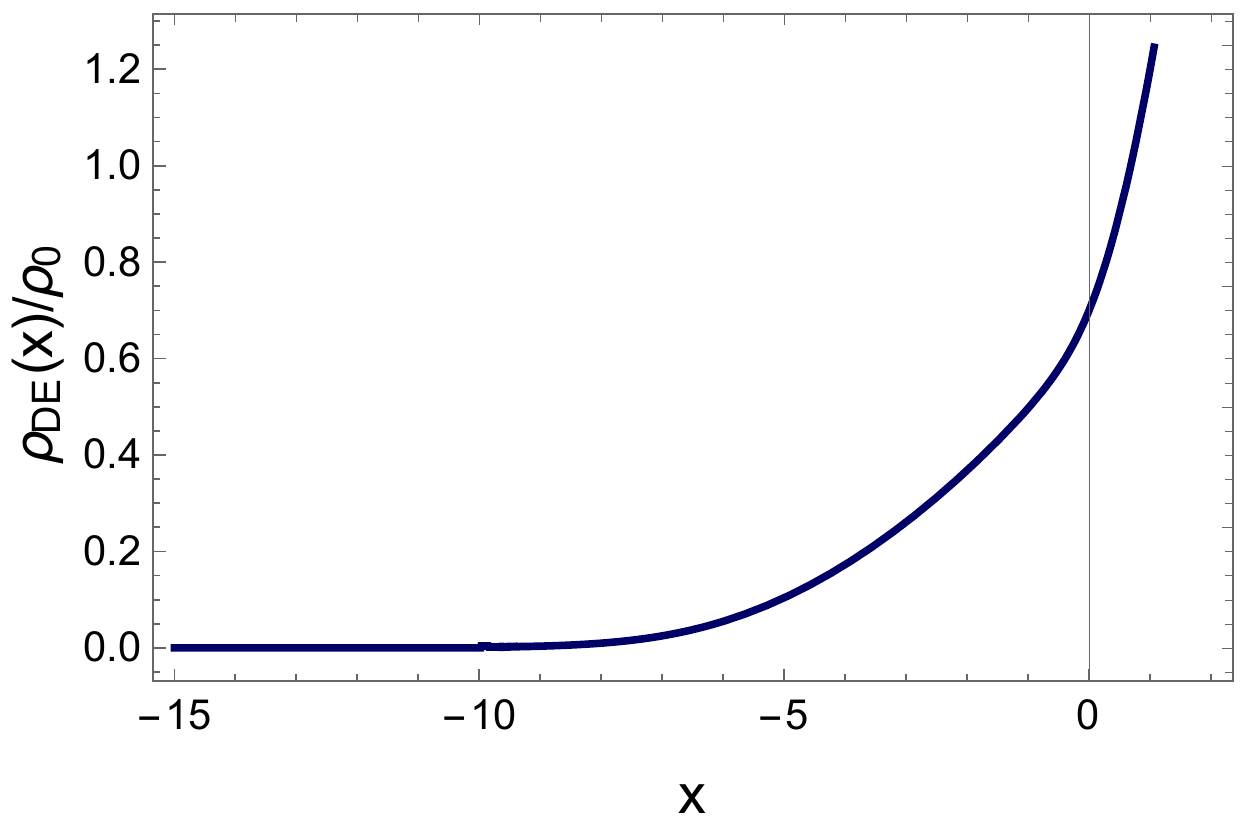}
\includegraphics[width=0.45\columnwidth]{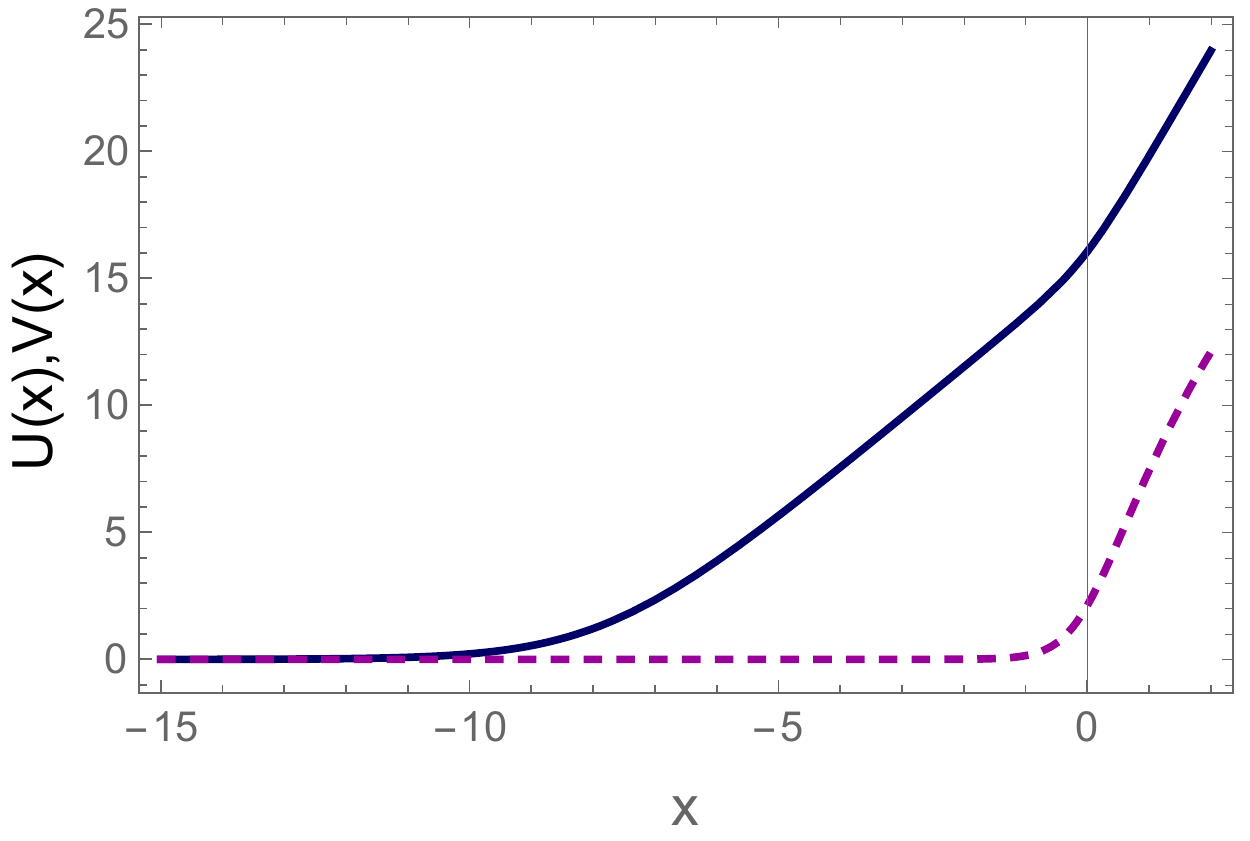}
\caption{\label{fig:RRu0zero} Left panel: the function $\rde(x)/\rho_0$,  against $x\equiv \ln a$, for the RR model with $u_0=0$.
Right panel:   the background evolution of the auxiliary fields $U$  (blue solid line) and $V$ (magenta dashed line).
Adapted from \cite{Dirian:2014ara}.
}
\end{figure}

\begin{figure}[t]
\centering
\includegraphics[width=0.45\columnwidth]{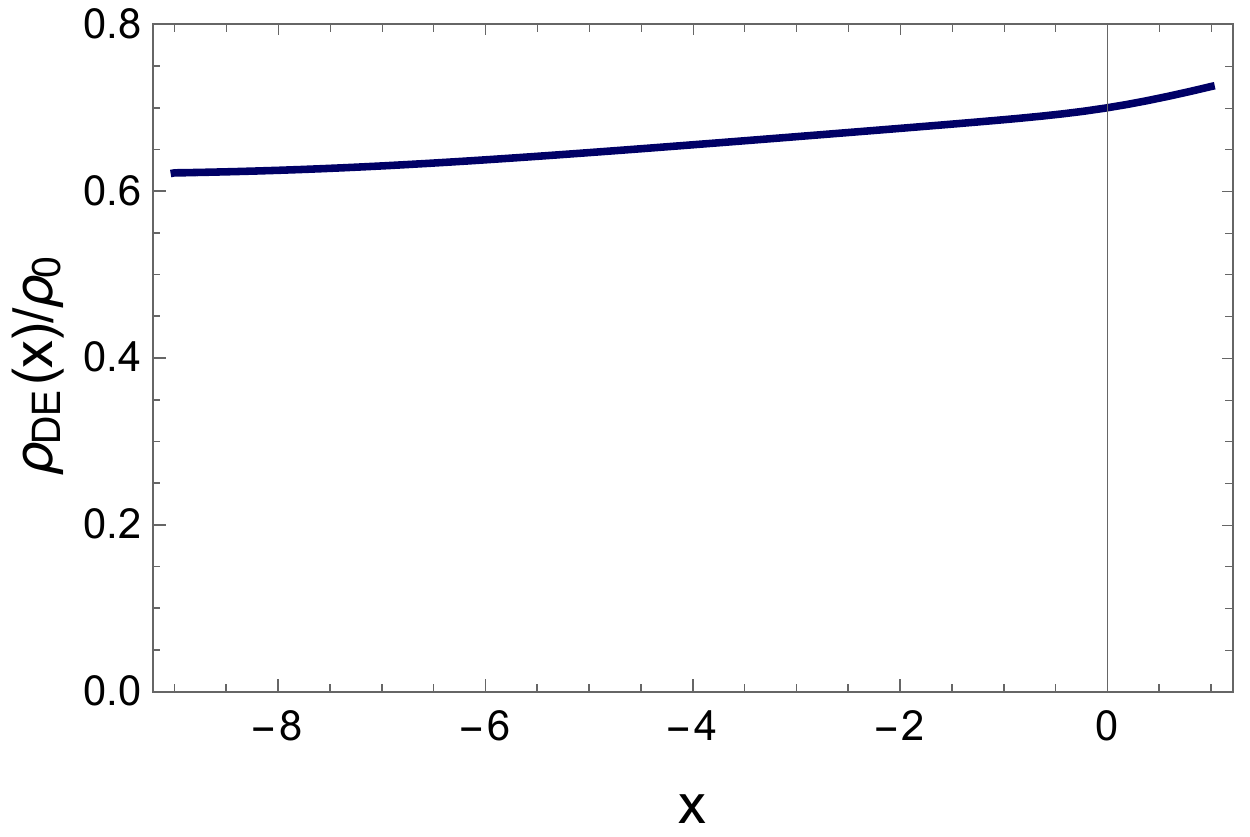}
\includegraphics[width=0.45\columnwidth]{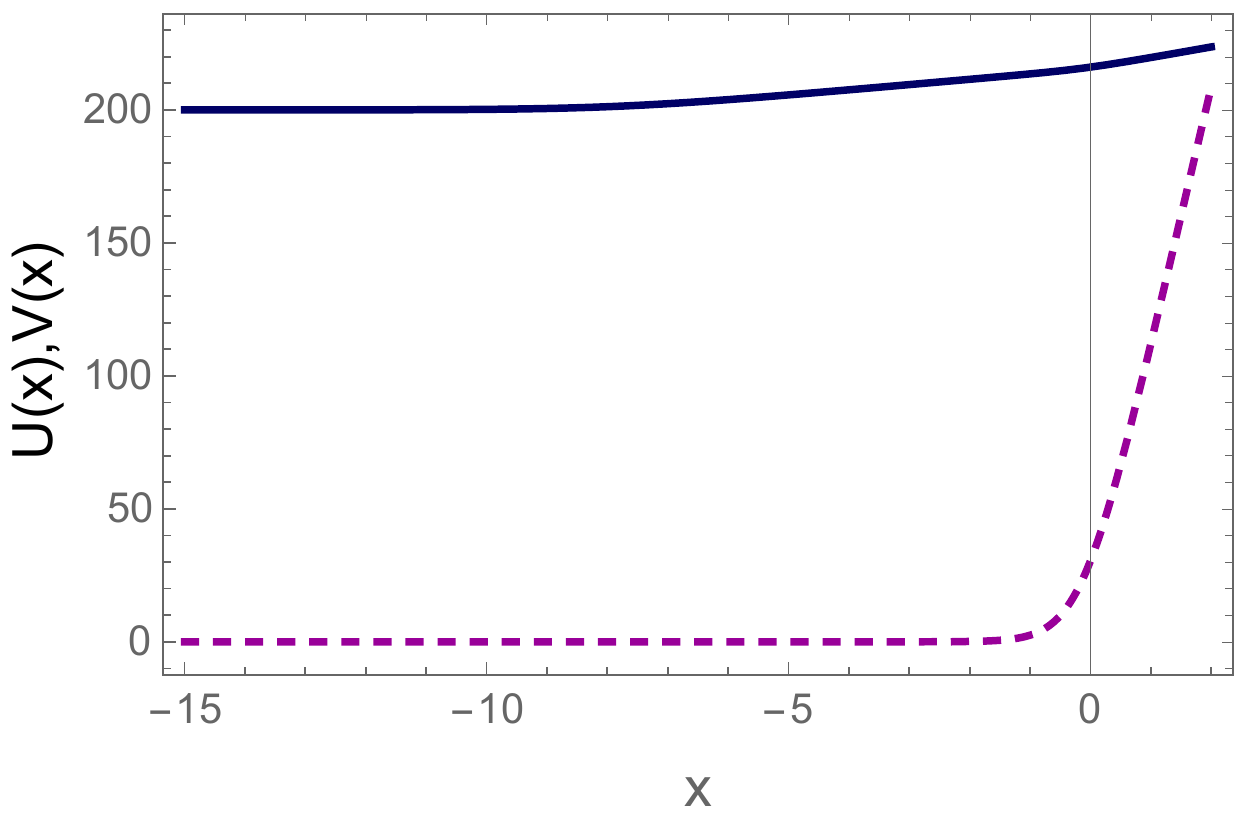}
\caption{\label{fig:RRu0_200} As in Fig.~\ref{fig:RRu0zero}, for $u_0=200$.
}
\end{figure}

\begin{figure}[t]
\centering
\includegraphics[width=0.45\columnwidth]{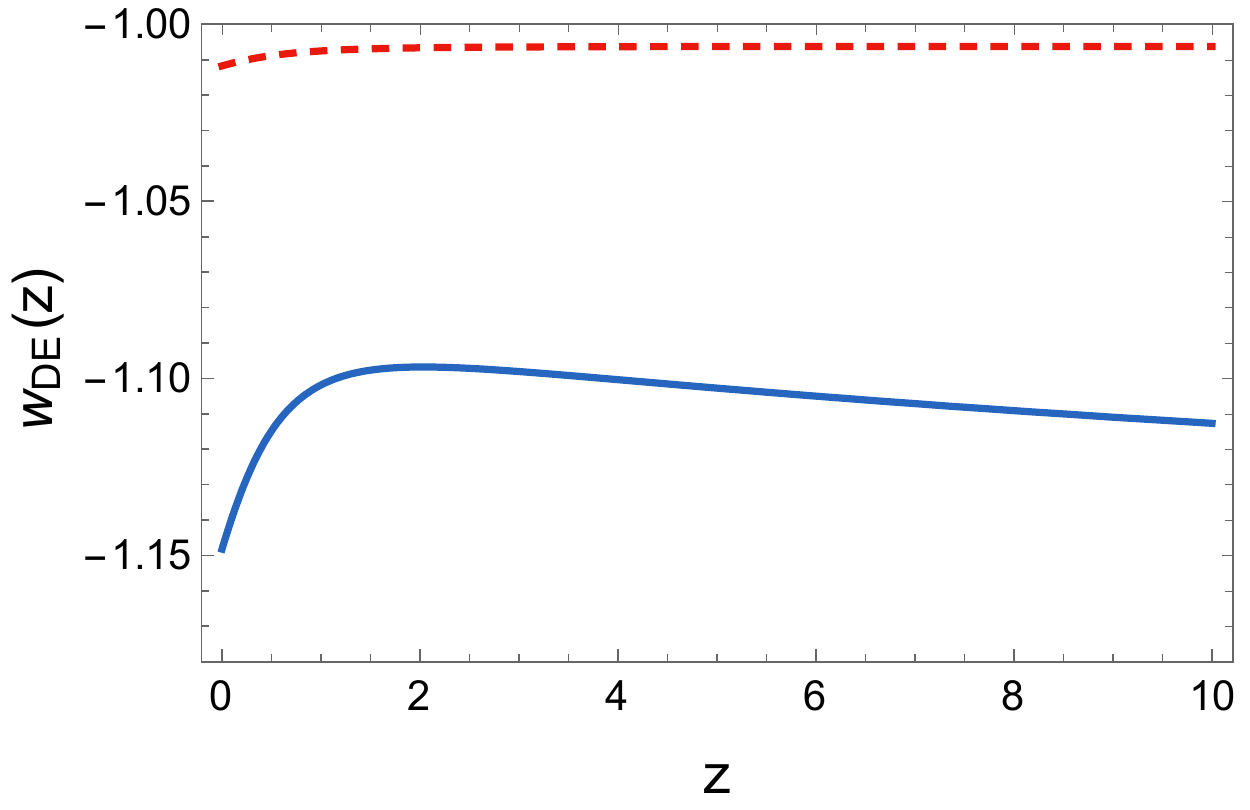}
\caption{\label{fig:wDE_RR_u0_0e200} The DE equation of state $\wde(z)$ as a function of redshift, for $u_0=0$ (blue solid line) and  $u_0=200$ (red dashed line).
}
\end{figure}

The result of the numerical integration is shown in Fig.~\ref{fig:RRu0zero} for $u_0=0$ and in
Fig.~\ref{fig:RRu0_200} for $u_0=200$. For illustration, in both cases  the parameter $\gamma$ is adjusted so to get a value $\rde/\rho_0=0.7$ today, which is obtained by setting
$\gamma\simeq 9.126\times 10^{-3}$ for $u_0=0$ and $\gamma\simeq 5.907\times 10^{-5}$ for $u_0=200$.\footnote{Of course, for each value of $u_0$, the exact value of $\rde/\rho_0$, and therefore $\gamma$, must eventually be  fixed  by comparing the model with a set of cosmological data and performing Bayesian parameter estimation, see the analysis in \cite{Dirian:2014bma,Dirian:2016puz,Dirian:2017pwp}.}
In terms of the variable $x=\ln a$, radiation-matter equilibrium corresponds to $x\simeq -8.1$, while the present epoch corresponds to $x=0$. From the left panel in Fig.~\ref{fig:RRu0zero} we see that, if we set $u_0=0$, the auxiliary fields, and therefore $\rde$, start from zero during RD, and then grow rather steeply during MD and the present DE-dominated era. This is due to the fact that $R=0$ in RD, so with the initial condition $u_0=0$ we have  $\bar{U}(x)=0$. As we enter in MD, $R$ becomes non-zero and drives the growth of $\bar{U}(x)$. In turn, $\bar{U}(x)$ will eventually drive the growth of $\bar{V}(x)$, through \eq{syV}. We see indeed from the right panel in Fig.~\ref{fig:RRu0zero} that the growth of $\bar{V}(x)$ is delayed compared to the growth of $\bar{U}(x)$. 
For large $u_0$, we see from Fig.~\ref{fig:RRu0_200} that $\bar{U}(x)$ [and therefore $\rde(x)$] is non-vanishing already in RD, and it evolves much more slowly. The evolution of the DE density, and therefore the whole cosmological evolution, is then much closer to that of $\Lambda$CDM. This can also be quantified in terms of the DE equation of state $\wde(x)$, defined as usual from the conservation equation
${\rho}'_{\rm DE}+3(1+w_{\rm DE})\rho_{\rm DE}=0$. Observe that, since both $\rde(x)$ and its derivative ${\rho}'_{\rm DE}=d\rde/dx$ are positive, necessarily $[1+\wde(x)]$ is negative, i.e. the DE equation of state is on the phantom side. Fig.~\ref{fig:wDE_RR_u0_0e200} shows $\wde$ as a function of
redshift $z$ [related to $x=\ln a $ by $x=-\log (1+z)$] for $u_0=0$ and $u_0=200$. We see that, for $u_0=200$, $\wde(z)$ differs from the $\Lambda$CDM value $-1$ only by about $1\%$, with 
$\wde(z=0)\simeq -1.01$, while for $u_0=0$ the deviation from $\Lambda$CDM  is much larger, and
$\wde(z=0)\simeq -1.15$.

\subsection{Static solution for the RR model}\label{sect:staticRR}

We next study the static \Sch solution for the RR model,  expanding  the analysis performed in \cite{Kehagias:2014sda,Maggiore:2014sia}. In this subsection, for notational simplicity, we denote simply by $U(r)$ and $S(r)$ the solutions denoted above as
$U_{\rm static}(r)$ and 
$S_{\rm static}(r)$. We consider a static spherically-symmetric metric of the form
\be\label{ds2}
ds^2=-A(r) dt^2+B(r)dr^2 +r^2(d\theta^2+\sin^2\theta\,  d\phi^2)\, .
\ee
Using the notation $A(r)=e^{2\a(r)}$ and
$B(r)=e^{2\b(r)}$ and
proceeding as in GR,  we get two independent equations for $\a(r)$ and $\b(r)$   by taking the combinations 
$e^{2(\b-\a)}R_{00}+R_{11}$ and $R_{22}$. 
In the RR model, using \eq{Gmn},  this gives
\bees
&&\hspace*{-15mm}\(1-\frac{m^2}{3} S\)(\pa_r\a+\pa_r\b)=\frac{m^2}{6} r \[(\pa_r\a+\pa_r\b-\pa_rU)\pa_r S-\pa_r^2S\]\, ,\label{eqab1}\\
&&\hspace*{-15mm}\(1-\frac{m^2}{3} S\)\left\{1+e^{-2\b}\[r(\pa_r\b-\pa_r\a)-1\]\right\}=\frac{m^2}{6} \[ r^2\(U+\frac{1}{2}U^2\)-2 r e^{-2\b}\pa_r S\]\, ,\label{eqab2}
\ees
which reduce to their GR counterparts for $m=0$. In the metric (\ref{ds2}) the equations for the auxiliary fields, \eqs{BoxURR}{BoxS}, become
\bees
&&r^2\pa_r^2U+[2r+(\pa_r\a-\pa_r\b)r^2]\pa_rU
=-2 e^{2\b}\nn\\
&&\hspace*{30mm}+2\[1+2r(\pa_r\a-\pa_r\b)+r^2\(\pa_r^2\a+(\pa_r \a)^2-\pa_r\a \pa_r\b\)\]\, ,
\label{eqU}\\
&&\pa_r^2S +\(\pa_r\a-\pa_r\b +\frac{2}{r}\) \pa_rS=-e^{2\b}U\, .\label{eqS}
\ees
The solution of these equations has been briefly discussed in \cite{Maggiore:2014sia},  following the detailed analysis for the RT model performed in \cite{Kehagias:2014sda}. In app.~\ref{App:Sch} we give a more detailed derivation specific to the RR model, while here we just give the results. 
We are only interested  in the solution at $r\gg r_S\equiv 2GM$, which is the regime relevant for studying solar system tests or LLR.  
For the metric, the solution for $r\gg r_S$ (and $mr$ generic) is  the same as for the RT model,
\bees
\hspace*{-2mm}A(r)&=&1-\frac{r_S}{r}\[1+\frac{1}{3}(1-\cos mr)\]\, ,\label{NewtA}\\
\hspace*{-2mm}B(r)&=&1+\frac{r_S}{r}\[1-\frac{1}{3}(1-\cos mr-mr \sin mr)
\]\, .\label{NewtB}
\ees
In particular, for $r_S\ll r\ll m^{-1}$ we have
\be\label{Afinal}
A(r)\simeq 1-\frac{r_S}{r}\(1+\frac{m^2 r^2}{6}\)\, ,
\ee
and $B(r)\simeq 1/A(r)$. Observe that this reduces smoothly to the \Sch solution of GR in the limit $m\ra 0$, contrary  to what happens in massive gravity where, at the linearized level, there is the  vDVZ discontinuity. The solution for the auxiliary fields $U(r)$ and $S(r)$, again in the limit $r\gg r_S$, is 
\bees
U(r)&=&\frac{r_S}{r}\cos (mr) \, .\label{NewtU}\\
m^2 S(r)&=&-\frac{r_S}{r}\[ 1-\cos(mr)\]\, .\label{solm2Sfinal}
\ees
In the region $r_S\ll r\ll m^{-1}$ these solutions  reduce to
\be\label{NewtUrmsmall}
U(r)\simeq \frac{r_S}{r} \, , \qquad m^2 S(r)\simeq -\frac{r_S}{2r}\, (mr)^2\, .
\ee
Note that the field $S$ has dimensions of inverse mass squared, so the combination $m^2S$  in \eq{solm2Sfinal} is dimensionless, and is the one that enters the effective Newton constant (\ref{Geff1LLR}). Observe that, through \eq{Geff1LLR}, a spatial dependence of $S$ induces a spatial dependence of Newton's constant. 
However, the resulting modification of Newton's constant must be contained in the motion of a 
test particle in the \Sch metric (\ref{ds2}), (\ref{NewtA}), (\ref{NewtB}). The latter shows that any deviation from GR vanishes in the limit $mr\ra 0$. This is indeed in agreement with \eq{solm2Sfinal}, that shows that, in the static context that we are considering here (i.e. a \Sch solution that at infinity reduces to flat Minkowski space, rather than being matched to FRW),  in the region  $r_S\ll r\ll m^{-1}$ we have 
\be\label{Geffstatic}
\frac{G_{\rm eff}(r)}{G} \equiv  \[1-\frac{1}{3}m^2 S(r)\]^{-1}
\simeq 1-\frac{1}{6}\, \frac{r_S}{r} (mr)^2\, ,
\ee
so the deviations are suppressed not only by $r_S/r\ll 1$ but also by a further factor $(mr)^2$ which, for $m\sim H_0$ (as determined by the cosmological solution) and $r$ of order, say, of the solar system scale, is totally negligible.

\subsection{A simple model for the matching: McVittie's metric}\label{sect:McV}

In the two previous subsections we have studied separately the cosmological solution of the RR model, in which the source is just given by the cosmic fluid, and the static solution, in which the source is just a static central mass and the metric at infinity approaches the flat Minkowski metric. We now wish to put together these solutions, to understand what is the short-distance solution for the  auxiliary field $S$ that determines the effective Newton constant,  when at large distances the metric approaches FRW. We begin with a somewhat idealized but still useful setting, which is provided by the McVittie metric~\cite{McVittie1933} (see e.g. \cite{Carrera:2008pi} for review). This metric can be taken as a model of a spacetime
sourced by a central mass $M$ plus a cosmic fluid, at distances much greater than the \Sch radius of the central mass. To introduce it let us recall first of all that, if we denote by $\vx$ the comoving coordinates in FRW and by ${\bf r}=a(t)\vx$ the physical coordinates, the flat FRW metric can be rewritten as
\be\label{FRWphys}
ds^2=-\[1-r^2 H(t)\] dt^2-2r H(t) dr dt +dr^2+r^2d\Omega^2\, ,
\ee
where $r=|{\bf r}|$. Physical coordinates are particularly useful for matching this solution to the \Sch metric generated by a central mass $M$. Indeed, in physical coordinates, the McVittie metric takes the very simple form~\cite{Nandra:2011ug}
\be\label{McVmetric}
ds^2=-\[1-\frac{r_S}{r}-r^2 H^2(t)\] dt^2-2r H(t) \( 1-\frac{r_S}{r}\)^{-1/2}dr dt +
 \( 1-\frac{r_S}{r}\)^{-1}dr^2+r^2d\Omega^2\, ,
\ee
where $r_S=2GM$. This metric reduces to the FRW solution (\ref{FRWphys}) as $M\ra 0$, and to the \Sch solution as $H\ra 0$. Observe also that, in $g_{00}$, the relative weight of the two terms $r_S/r$ and $H^2r^2$ is determined by the combination $r_c(t)$, defined by 
$r_S/r_c(t)=r_c^2(t) H^2(t)$, i.e.
\be
r_c(t)=\[ r_S H^{-2}(t)\]^{1/3}\, .
\ee
For $r\gg r_c$ we are in the cosmological regime, where the FRW term dominates, while for $r\ll r_c$ the metric is dominated by the central mass.

Note however that, unlike the \Sch metric, which in $r=r_S$ only has a coordinate singularity,  the McVittie metric has a genuine  physical singularity at $r=r_S$. Indeed, if one computes the energy-momentum tensor that sources this metric, one finds a fluid with a pressure $p(t,r)$ that, at $r\gg r_S$, correctly reduces to the cosmological fluid corresponding to the chosen time dependence of $H(t)$, but that diverges as $r\ra r_S$. For instance, when  matching to a matter-dominated universe at cosmological scales, one finds that the relation between pressure and energy density is~\cite{Nandra:2011ug}
\be\label{pMcV}
p(t,r)=\rho(t)\[  \( 1-\frac{r_S}{r}\)^{-1/2} -1 \]\, .
\ee 
This correctly reduces to the result $p=0$ for non-relativistic matter at $r\gg r_S$, but diverges as $r\ra r_S$. Thus, the  McVittie metric becomes pathological, and is no longer a useful model for a mass embedded in an expanding universe, when $r$ is close to $r_S$. However, this is only relevant if we want to describe black holes or compact objects in an expanding universe. For a central object of size much larger than its \Sch radius, the metric in the exterior region can be very well approximated by \eq{McVmetric}, and the pressure $p(t,r\gg r_S)$ is an excellent approximation to that of the desired cosmic fluid.

Thus, we can use the McVittie metric as a first simple toy model to study the effect of the cosmic expansion on the Earth-Moon system (with the Earth playing the role of the central mass $M$ and the Moon of a test mass), and in particular to study whether, in this metric, the field $S$ of the RR model, that determines the effective Newton constant, retains a time dependence at small distances. Of course, the McVittie metric is a solution of the Einstein equations of GR, and not of those of the RR model (as it is clear from the fact that it does not even depend on the parameter $m$). This means that, when $mr$ is no longer negligible, in the RR model we should expect deviations from it. In this section we will then consider a slightly more general metric
\bees\label{Phimetric}
ds^2&=&-\[1-2\Phi(r)-r^2 H^2(t)\] dt^2-2r H(t) \[ 1-2\Phi(r)\]^{-1/2}dr dt \nn\\
&&+
 \[ 1-2\Phi(r)\]^{-1}dr^2+r^2d\Omega^2\, ,
\ees
with a more general function $\Phi(r)$. The McVittie metric  corresponds to $2\Phi(r)=r_S/r$. The simplest possibility is that, in the RR model, the function $\Phi(r)$ will receive corrections  $1+{\cal O}(m^2r^2)$. More generally, the various occurrences of $\Phi$ in \eq{Phimetric} could be modified in different ways as $mr$ becomes non-negligible. However, as a first toy model, \eq{Phimetric} will already be sufficient to understand the behavior of the solutions of 
the equations $\Box U=-R$ and $\Box S=-U$ on a background that interpolates between a \Sch metric at small scales and FRW at large scales. Furthermore, from the analysis in the similar setting  performed in app.~\ref{App:Sch}, we understand that $\Phi(r)$ will be suppressed by a factor $r_S/r$ even at cosmological scales, i.e. when $mr$ is not small, see 
\eq{PhicPhi} and is therefore extremely small at cosmological scales. This also means that we do not need  knowledge of the exact way in which the solution of the RR model differs from the McVittie solution of GR as $mr$ becomes of order  one.

Actually, \eq{Phimetric} must be understood as a first step in an iterative procedure in which $\Phi(r)$ is more generally replaced by a function that also has an adiabatic dependence on time, $\Phi=\Phi(t,r)$. Indeed, if the analysis in the metric (\ref{Phimetric})  should lead to the conclusion that $S$ depends on time even at short scales, as indeed we will show below, it would follows that the dynamics at short distances is governed by $G_{\rm eff}(t)$, rather than by $G$. In turn, this means that the occurrencies of $G$ that are implicitly present in $\Phi(r)$, e.g. the factor $G$ in $\Phi(r)=r_S/(2r)=GM/r$, must be replaced by $G_{\rm eff}(t)$, and correspondingly $\Phi(r)$ becomes a function $\Phi(t,r)$ with a  dependence on time. Since in any case 
$G_{\rm eff}(t)/G$ is very close to one, an iterative approach of this kind, in which we start at first with a function $\Phi(r)$ in \eq{Phimetric}, is appropriate. We will see explicitly how this iterative procedure works in the discussion below \eq{BoxUansatz}.

One might argue that this setting is  however still oversimplified because, to move from  the Earth-Moon system scale to cosmological scales we have to go through a hierarchy of embeddings, from  Solar System to the Galaxy, Local Group, Clusters and Superclusters, and one might wonder whether each such embedding could  provide some form of screening, so that the matching of the local \Sch metric to the cosmological FRW solution will be more complicated~\cite{Carrera:2008pi}. We will address this question in section~\ref{sect:pertFRW} and, for the moment, we  just proceed with the exercise in the McVittie-like  metric (\ref{Phimetric}).  The use of a generic function $\Phi(r)$, rather than the explicit expression $r_S/(2r)$ corresponding to the McVittie solution of GR, will also allow us to have a better understanding of the structure of the equations, and to more easily make  contact 
with the analysis of section~\ref{sect:pertFRW}.

We will assume that, in the region of interest (such as $r\gg r_S$ for a single central mass with \Sch radius $r_S$),  we have
\be\label{condPhi1}
|\Phi(r)|\ll 1\, ,
\ee
and also 
\be\label{condPhi2}
|r\pa_r\Phi|\ll 1\, .
\ee 
This is the case in particular for the McVittie metric when $r\gg r_S$, since in that case $\Phi(r)=r_S/(2r)$ and
$r\pa_r\Phi(r)=-r_S/(2r)$.  
In the metric (\ref{Phimetric}),  neglecting terms ${\cal O}(\Phi^2)$, the equations
$\Box U=-R$ and $\Box S=-U$ become
\bees
&&(1-2\Phi-H^2r^2)\(\pa_r^2U+\frac{2}{r}\pa_rU\)-\[2r\pa_r\Phi+r^2(\dot{H}+2H^2)\]\frac{1}{r}\pa_rU
-2rH\pa_t\pa_rU\nn\\
&&-(1+2\Phi)\pa_t^2U-3H \(1+\Phi+\frac{1}{3}r\pa_r\Phi\)\pa_tU=-R\, ,\label{BoxUMcV}
\ees
(where $\dot{H}=\pa_tH$), and
\bees
&&(1-2\Phi-H^2r^2)\(\pa_r^2S+\frac{2}{r}\pa_rS\)-\[2r\pa_r\Phi+r^2(\dot{H}+2H^2)\]\frac{1}{r}\pa_rS
-2rH\pa_t\pa_rS\nn\\
&&-(1+2\Phi)\pa_t^2S-3H \(1+\Phi+\frac{1}{3}r\pa_r\Phi\)\pa_tS=-U\, .\label{BoxSMcV}
\ees
In the RR model the Ricci scalar $R$, that appears on the right-hand side of \eq{BoxUMcV}, can be determined by taking the trace of \eq{Gmn},
 which gives
\be\label{Rfromtrace}
\(1-\frac{1}{3}m^2 S\)R=-8\pi G T +\frac{1}{3}m^2  (3U+U^2-\pam S\paM U)\, .
\ee
For a single fluid $T=-\rho+3p$, so for a cosmological fluid with energy density and pressures
$\rho_{\rm cosmo}(t)$ and $p_{\rm cosmo}(t)$, respectively, plus a central mass $M$,
\be\label{Ttrace}
T=-\rho_{\rm cosmo}(t)+3p_{\rm cosmo}(t)-M\d^{(3)}({\bf r})\, .
\ee
We  now study whether \eqs{BoxUMcV}{BoxSMcV} have an approximate solution
\be\label{ansatz1}
U(t,r)\simeq U_{\rm cosmo}(t)+U_{\rm static}(r)\, ,\quad
S(t,r)\simeq S_{\rm cosmo}(t)+S_{\rm static}(r)\, ,
\ee
where $U_{\rm cosmo}(t)$ is the cosmological solution discussed in section~\ref{sect:cosmoRR} and
$U_{\rm static}(r)$ is the static \Sch solution of section~\ref{sect:staticRR}. The details of the analysis are given in App.~\ref{app:McV}, while here we give the main results.

On the left-hand side of \eq{Rfromtrace} we can make use of the fact that $m^2S_{\rm static}(r)$ is always very small compared to one, as long as $r\gg r_S$, as we see from \eq{solm2Sfinal}, so we can replace
$[ 1-(1/3)m^2 S]$ by $[ 1-(1/3)m^2 S_{\rm cosmo}(t)]$.
We then introduce the notation
\bees
R_{\rm static}(r)
&=&8\pi G\, M\d^{(3)}({\bf r})   +\frac{m^2}{3}\, \[  3U_{\rm static}+U^2_{\rm static}-g^{rr}\pa_r S_{\rm static}\pa_r U_{\rm static}\] \, ,\\
R_{\rm cosmo}(t)&=& 8\pi G\, \frac{ \rho_{\rm cosmo}(t)-3p_{\rm cosmo}(t)  }{1-(m^2/3)S_{\rm cosmo}(t)  }\nn\\
&& +\frac{m^2}{3}\, \frac{  3U_{\rm cosmo}+U^2_{\rm cosmo}-g^{00}\pa_t S_{\rm cosmo}\pa_t U_{\rm cosmo}}{1-(m^2/3)S_{\rm cosmo}(t) }\, .
\ees
Then, the analysis in App.~\ref{app:McV} shows that,
on the ansatz (\ref{ansatz1}), the equation $\Box U=-R$ takes the form
\be\label{BoxUansatz}
 -\(\pa_t^2\Uc+3H(t)\pa_t\Uc\)+\(\pa_r^2U_{\rm static}+\frac{2}{r}\pa_rU_{\rm static}\)
\stackrel{?}{=}-R_{\rm cosmo}(t)-\frac{G_{\rm eff}(t)}{G} R_{\rm static}(r) \, ,
\ee
where the question mark stresses that we are just checking if the ansatz is correct, and 
\be\label{GeffsuGstatcosm}
\frac{G_{\rm eff}(t)}{G}=\frac{1}{1-(1/3)m^2 S_{\rm cosmo}(t) }\, .
\ee
By definition, $U_{\rm cosmo}(t)$ and $S_{\rm cosmo}(t)$ solve the equation
\be
-\(\pa_t^2\Uc+3H(t)\pa_t\Uc\)=-R_{\rm cosmo}(t)\, ,
\ee
while  $U_{\rm static}(r)$ and $S_{\rm static}(r)$ solve 
\be
\(\pa_r^2U_{\rm static}+\frac{2}{r}\pa_rU_{\rm static}\)=- R_{\rm static}(r)\, .
\ee
Comparing with \eq{BoxUansatz} we see that the ansatz (\ref{ansatz1}) would indeed be a solution of the equations, were it not for the term $G_{\rm eff}(t)/G$
on the right-hand side of \eq{BoxUansatz}. Using, as in section~\ref{sect:cosmoRR}, $\gamma=m^2/(9H_0^2)$ and $V(t)=H_0^2S_{\rm cosmo}(t)$, we have 
\be
\frac{1}{3}m^2S_{\rm cosmo}(t)=3\gamma V(t)\, .
\ee 
Parametrically, this is not suppressed by any large factor.  However, from the numerical solution of the cosmological equations discussed in section~\ref{sect:cosmoRR}, we know  that $G_{\rm eff}(t)/G$  is always very close to one, in the past and present cosmological evolution. For instance, for $u_0=0$ we have $\gamma\simeq 9.126\times 10^{-3}$ and $V(t_0)\simeq 2.14$, so
$3\gamma V(t_0)\simeq 0.06$. For large values of $u_0$ this quantity decreases further.
For instance, for $u_0=200$ we have  $\gamma\simeq 5.907\times 10^{-5}$, $V(t_0)\simeq 30.45$ and 
$3\gamma V(t_0)\simeq 0.005$. 
Thus, we can devise a perturbative strategy in the small quantity
$3\gamma V(t)$. To zero-th order we just neglect it, and then $G_{\rm eff}(t)=G$. Of course, to this order we have lost the effect that we are looking for, namely whether the effective Newton constant is time dependent. We must then go to the next order. To this purpose we start from the expression
\be\label{GeffsuGS0}
\frac{G_{\rm eff}}{G}=\(1-\frac{m^2}{3}S\)^{-1}\, ,
\ee 
and then, to compute $S$ on the right-hand side, we  set $G_{\rm eff}=G$ in \eq{BoxUansatz}. In this case, we see that the ansatz (\ref{ansatz1})  indeed  solves (within the approximations discussed)  $\Box U=-R$, and precisely the same analysis applied to $\Box S=-U$ shows that indeed $U=U_{\rm cosmo}(t)+U_{\rm static}(r)$ and $S=S_{\rm cosmo}(t)+S_{\rm static}(r)$ is an approximate solution of both $\Box U=-R$ and $\Box S=-U$.  Plugging this solution for $S$ in \eq{GeffsuGS0} we get $G_{\rm eff}/G$ to first order in $3\gamma V(t)$,
\be\label{GeffsuGS1}
\frac{G_{\rm eff}(t)}{G}=\[1-\frac{m^2}{3}S_{\rm cosmo}(t)\]^{-1}\, ,
\ee 
which is indeed the  expression used in \cite{Barreira:2014kra} and in section~\ref{sect:LRRvsRR}. We now see more formally how this emerges, at least within the McVittie model, and under what approximations.

One could in principle iterate the process. It is clear that, at next order, the result will basically amount to replacing the static solutions $U_{\rm static}(r)$ and $S_{\rm static}(r)$ with a quasi-static solution that is obtained by replacing $G$ with the $G_{\rm eff}(t)$ computed at first order. The physics behind this iterative procedure is quite clear. If Newton's constant has a slow time dependence on cosmological time-scales, on the time-scale  ${\cal O}(10)$~yr  probed by LLR the change in the Newton constant has the effect of an adiabatic perturbation, and the corresponding \Sch solutions at $r\gg r_S$ is simply obtained by replacing $G$ with  $G_{\rm eff}(t)$.

The conclusion of this section is that, within the modelisation of the problem provided by the McVittie solution, it is indeed true that, even at the Earth-Moon scale  probed by LLR, Newton's constant is replaced by an effective time-dependent Newton's constant, just as on cosmological scales. To a precision of order of a few percent or better, the corresponding expression is given by \eq{GeffsuGS1}.\footnote{Indeed, the accuracy is probably much better since, at next iterative order, we expect that the only change would be given by the replacement of 
$U_{\rm static}(r)$ and $S_{\rm static}(r)$ with the corresponding quasi-static solutions. Since, in any case, in $S=S_{\rm cosmic}(t)+S_{\rm static}(r)$ we have seen that $S_{\rm cosmic}(t)$ is numerically much larger than $S_{\rm static}(r)$, the result of successive iterations will have very little effect on
$G_{\rm eff}(t)/G$.} As we found in section~\ref{sect:LRRvsRR}, in this case the LLR bound is violated by a factor that, for $u_0=0$, is ${\cal O}(100)$, and even for $u_0=200$ is still violated by about one order of magnitude, so the level of accuracy of the first-order solution (\ref{GeffsuGS1}) is perfectly adequate and shows that, within this McVittie  modelisation, the RR model is ruled out by LLR, unless we go to very large values of $u_0$, $u_0\, \gsim\,  {\cal O}(2000)$.

\subsection{A more general analysis: perturbed FRW}\label{sect:pertFRW}

The next question is whether the result found above is specific to our modelization of the problem. After all the McVittie metric, describing a single central mass in an expanding FRW universe, might look like a very simplified setting, compared to the actual setting in which the mass $M$ (the Earth for LLR) is embedded into a hierarchy of structures such as the solar system, the Galaxy, the Local Group, clusters and superclusters, before reaching the scales where the Universe can be approximated by a uniform FRW background. Before drawing our conclusions from a computation performed in the McVitties metric,
we should investigate whether the presence of these non-linear structures can somehow provide a screening mechanism. This requires to study the equations $\Box U=-R$ and $\Box S=-U$  on a metric that gives a realistic description of the Universe from the Earth-Moon scale to the cosmological scales. 
The most natural choice is to consider a perturbed  FRW spacetime (\ref{Poissongauge}).
In the absence of anisotropic stress\footnote{In the RR model the nonlocal term generates a small anisotropic stress, which however is completely negligible numerically, see Fig.~8 of ref.~\cite{Dirian:2014ara}.} we have $\Psi=-\Phi$, so
\be\label{Poissongauge}
ds^2=-(1-2\Phi) dt^2+(1+2\Phi)a^2(t)d\vx^2\, .
\ee
To make contact with the analysis of the previous subsection one could also re-express this metric using physical coordinates ${\bf r}=a(t)\vx$, which leads to
\be\label{pertFRWphys}
ds^2=-\[1-2\Phi -(1+2\Phi)H^2r^2\] dt^2 -2Hr (1+2\Phi) dr dt
+(1+2\Phi) (dr^2+r^2d\Omega^2)\, .
\ee
For McVittie metric, physical coordinates were particularly transparent for showing how the \Sch metric  merges with the FRW metric. However, in this section we will use comoving coordinates, since the corresponding equations turn out to be somewhat simpler. 

In recent years there has been much discussion on whether nonlinear structures can imprint a significant backreaction on the metric, in particular in the attempt of explaining the accelerated expansion of the Universe  (see e.g.~\cite{Buchert:2011sx} for review). The issue is controversial, but there is by now a broad consensus on the fact that no large backreaction appears if one uses the Newtonian  gauge (\ref{Poissongauge0}), both from analytic arguments (see e.g. \cite{Ishibashi:2005sj,Green:2014aga}) and from  $N$-body simulations~\cite{Adamek:2014gva,Adamek:2017mzb}. The metric  (\ref{Poissongauge0})  [or, neglecting anisotropic stresses, the metric
(\ref{Poissongauge})] gives an excellent approximation to the actual metric of the Universe (in the scalar-perturbation sector), and $|\Phi|$ never exceeds a value of order $10^{-4}$, except in strong-field regions, such as the vicinity of black holes or neutron stars. In particular, for a virialized system with characteristic velocity $v$, the virial theorem gives $\Phi\sim v^2$. However, the spatial derivatives  are enhanced. For the first derivative, from the Euler equation we get $\pa_i\Phi\sim {\cal H}v$ [where ${\cal H}=(1/a)\pa_{\eta}a$, $\eta$ is conformal time, and $\pa_i$ is the derivative with respect to comoving coordinates] so, in the natural units provided by the Hubble parameter, $\pa_i\Phi$ is suppressed only by one power of $v$. Furthermore, from Poisson equation, we have $\n^2\Phi\sim {\cal H}^2 (\delta\rho/\rho)$. In non-linear structures the overdensity $\delta\rho/\rho$ can be extremely large, say
${\cal O}(10^{30})$. Still, even in these cases $|\Phi|$ is small and \eq{Poissongauge} provides a good description of the metric. We will then take  the perturbed FRW metric (\ref{Poissongauge}) as a good description from cosmological scales down to the scale of the Earth-Moon system probed by LLR.

Of course, on the length-scales   and time-scales relevant  for LLR,  the effect of the explicit occurrence of $a(t)$ in \eq{Poissongauge} [or, equivalently, the effect of the $Hr$ and 
$H^2r^2$ terms in \eq{pertFRWphys}] is in any case completely negligible compared to the effect of the static or quasi-static potential $\Phi$ generated by the local mass distribution and, to excellent accuracy, on these scales  we might as well use a perturbed Minkowski metric, setting $a(t)=1$. Furthermore, as we recalled in section~\ref{sect:staticRR}, in  the RR model there is no vDVZ discontinuity and the effects associated to the mass $m$ go smoothly to zero on scales $r$ such that  $mr\ll 1$ (in contrast, e.g. to massive gravity, where, at the linearized level, there is the vDVZ discontinuity).  Thus, one might be tempted to conclude that, in the RR model, the motion of a test mass (such as the Moon) in the field of the Earth is in practice indistinguishable from that in GR.
As pointed out in \cite{Barreira:2014kra}, this is not necessarily correct
if Newton's constant  (that implicitly enters the  metric through $\Phi$, e.g. through the expression $\Phi(r)=GM/r$ in the case of a single  massive body) retains a time dependence down to such scales, because the extraordinary accuracy of the LLR measurement allows us to detect a rate of change of Newton's constant comparable to $H_0$ [and in fact even three order of magnitude smaller, see 
\eq{dotGsuGH0}]. In other words, the potentially dangerous time dependence of the metric at short scales does not come from the explicit occurrences of $a(t)$ in \eq{Poissongauge},  or of $H(t)$ in \eq{pertFRWphys}, but rather from a possible  implicit time dependence that could be present in $\Phi$ through a time-dependent Newton constant.
In turn, whether  Newton's constant evolves with time depends on whether the auxiliary fields $U$ and $S$, solutions of $\Box U=-R$ and $\Box S=-U$,  retain a time dependence on these small scales, due to the fact that they are time dependent on cosmological scales.
As we have seen, this is precisely what happens in the McVittie metric. 

We then  use  a perturbed FRW metric  to study the solution of the equations $\Box U=-R$ and $\Box S=-U$.
As in the previous section, we give here the main results, deferring to App.~\ref{app:pertFRW} some technicalities.
To first order in $\Phi$, for a generic function $f$,  in the metric (\ref{Poissongauge}) we have
\be\label{Boxffirstorder}
\Box f = -(1+2\Phi) (\ddot{f} + 3 H \dot{f})  - 4 \dot{\Phi} \dot{f} + a^{-2} (1-2\Phi)  \nabla^2 f\, ,
\ee
where $\n$ is the gradient with respect to the comoving coordinates.
We could now deal with the equation $\Box U=-R$ just as we did for the McVittie metric, looking for an approximate solution displaying a separation of variables.\footnote{In section~
\ref{sect:McV} we have shown the existence of a solution with an approximate separation of variables  between cosmic time $t$ and physical coordinates ${\bf r}$. We could as well have used cosmic time and comoving coordinates $\vx$. Indeed, for a differentiable function $f(t,\vx)$ we have $(\pa f/\pa x^i)_t=a(\pa f/\pa r^i)_t$ and
\be\label{patparf}
\(\frac{\pa f}{\pa t}\)_x=\(\frac{\pa f}{\pa t}\)_r+Hr^i\(\frac{\pa f}{\pa r^i}\)_t\, .
\ee
Taking $f(t,r)=U_{\rm cosmo}(t)+U_{\rm static}(r)$, we have $(\pa f/\pa t)_r\sim HU_{\rm cosmo}$, and 
$U_{\rm cosmo}={\cal O}(1-10)$ in the recent cosmological epoch. In contrast, in a McVittie-like metric,
$(\pa f/\pa r)_t\sim (1/r)U_{\rm static}$  and $U_{\rm static}\sim r_S/r\ll 1$ (one can also express  this by observing that spatial derivatives vanish on the FRW background, so they pick an extra factor of  order $\Phi$ with respect to time derivatives, and $\Phi\sim r_S/r\ll 1$). Then,
the second term on the right-hand side of \eq{patparf} is negligible with respect to the first (and similarly for $S$).} However, in the spirit of the use of a perturbed FRW metric, it is more natural to look for a perturbative solution of the form
\begin{align}
U(t,\vx) = U_{c}(t) + \delta U (t,\vx) \label{pertans1}\, ,\\
S(t,\vx) = S_{c}(t) + \delta S(t,\vx) \label{pertans2}\, ,
\end{align} 
where $\delta U$ is a perturbation with respect to $U_{c}$ and similarly for $S$. 
[In this section, to keep the equations compact, we denote the cosmological background solutions $U_{\rm cosmo}(t)$ and $S_{\rm cosmo}(t)$ by $U_c(t)$ and 
$S_c(t)$, respectively]. Of course, we already know from the analysis in \cite{Nesseris:2014mea,Dirian:2014ara} that a perturbative approach is consistent at cosmological scales, where the perturbations of the matter density are small. Here however we wish to see if such a perturbative solution is still consistent everywhere, including the `short distance' scales where there can be large density perturbations, and therefore without linearizing with respect to the matter density contrast $\delta_M\equiv \delta\rho_M/\rho_M$, as in \cite{Nesseris:2014mea,Dirian:2014ara}. The key to this more general treatment is that, as we have discussed, $|\Phi|$ still remains small even when its Laplacian, which is related to the density perturbations, is large.

We begin by discussing the equation $\square U = -R$.  We drop the ${\cal O}(\Phi)$ corrections (as we did in the analysis for the McVittie metric),  and we use the perturbative ansatz (\ref{pertans1}), (\ref{pertans2}).
In addition, we assume a quasi-static approximation 
\begin{align}
|\delta  \dot{U} |&\ll a^{-1} |\vec{\nabla} \delta U|\, , \label{quasi1}\\
|\delta  \dot{S}  | &\ll a^{-1}|\vec{\nabla} \delta S|\, ,\label{quasi2}
\end{align} 
where $\delta  \dot{U}\equiv \pa_t(\delta U)$, $\delta  \dot{S}\equiv \pa_t(\delta S)$, and we also assume 
\begin{align}\label{eq:smallDeltaS}
|m^2 \delta S |\ll |\delta U |\, ,
\end{align}
that will be checked  a posteriori.
Using these  approximations, we show in App.~\ref{app:pertFRW} that the equation $\Box U=-R$ reduces to
\be\label{dUrhominusrhob}
-4 \dot{\Phi} \dot{U}_c + a^{-2}   \nabla^2 \delta U \simeq  - \frac{ 8 \pi G   }{1-m^2 S_c/ 3}  (\rho-\rho_b),
\ee
where $\rho_b$ is the background energy density of the cosmological fluid.
We now make use of the fact that, in the sub-horizon limit,  $\Phi$ satisfies the Poisson equation of the RR model,\footnote{The Poisson equation is a combination of the $00$ and of the divergence of $0i$ components of the  modified Einstein equation of the RR model, in the quasi-static and sub-horizon limit, and does not rely on the use of  linear cosmological perturbation theory. Therefore,
just as in GR, also for the RR model  it holds even for large density contrasts.}
\begin{align}
  -a^{-2} \nabla^2 \Phi = \frac{ 4 \pi G  }{1-m^2 S_c/ 3}  (\rho-\rho_b)\, .
\end{align} 
Thus, \eq{dUrhominusrhob} can be rewritten as
\be\label{nablePhiU}
 a^{-2}   \nabla^2 (\delta U -2\Phi)\simeq 4 \dot{\Phi} \dot{U}_c\, .
 \ee
 Next observe that, for modes with wavenumber $k$, the left-hand side is of order 
 \be
 \frac{k^2}{a^2}\, (\delta U_k -2\Phi_k)\, ,
 \ee 
 while, using the quasi-static approximation, 
 \be
| \dot{\Phi} \dot{U}_c|\ll 
a^{-1}  |\vec{\nabla} \Phi| \dot{U}_c\sim a^{-1}  |\vec{\nabla} \Phi| H U_c\, ,
\ee
and $U_c$ is ${\cal O}(1-10)$, so in momentum space
\be
| \dot{\Phi} \dot{U}_c|\ll \frac{k}{a} H |\Phi_k|\, .
 \ee
Inside the horizon $H\ll k/a$, so the right-hand side of \eq{nablePhiU} is parametrically smaller than the left-hand side, and we must have 
\be\label{deltaU2Phiinside}
\delta U\simeq 2\Phi\, .
\ee
Thus, the smallness of $\Phi$ ensures the  smallness of $\delta U$ with respect to $U_c$, which in contrast is ${\cal O}(1-10)$, and validates the use of the perturbative ansatz (\ref{pertans1}). 

To conclude our argument we need to check the consistency of the assumption~(\ref{eq:smallDeltaS}). 
This is obtained for a study of the equation $\Box S=-U$, and again the details are given in 
App.~\ref{app:pertFRW}.
Note that \eq{eq:smallDeltaS}  also implies that $ |\delta S| \ll |S_c|$, since $|\delta U| \ll  U_c \sim H_0^2 S_c \sim m^2 S_c $ today, and therefore confirms the validity of the perturbative approach also for $S$.

Therefore, the perturbative ansatz gives a consistent solution to the equations for the auxiliary fields in terms of a generic metric perturbation $\Phi$. In turn, the latter can be determined from the modified Einstein equation (\ref{Gmn}), providing a fully consistent perturbative solution of the equations of the RR model.
Furthermore,  the perturbative ansatz obeys sensible initial conditions, matching to the linear perturbation theory solution at early times. Indeed, before the epoch of non-linear structure formation $\rho$ is very close to $\rho_b$ and,  from \eq{dUrhominusrhob},  $\delta U$ is basically zero. In other words what we have found is that, before structure formation, when the homogeneous FRW metric holds to great accuracy everywhere, 
the fields $U$ and $S$ evolve with time according to the cosmological solution. More precisely, as we
recalled in section~\ref{sect:cosmoRR}, deep in RD the cosmological solutions for the auxiliary fields is basically constant in time, but the evolution of $U$ becomes non-negligible as we enter in MD, even before structures become non-linear.
As  structures form and become non-linear, the solutions for $U$ and $S$ remains a  small perturbations of the cosmological solution, of the form (\ref{pertans1}), (\ref{pertans2}), as a reflection of the fact that the metric perturbations never become large (except, of course, in the very limited regions around compact bodies). Thus, structure formation does not stop the time evolution of the auxiliary fields.

It is interesting to compare the results of this section with those obtained in section~\ref{sect:McV} in the McVittie metric. The result $\delta U=2 \Phi$ that we have found inside the horizon in the very  general setting of a perturbed FRW spacetime is confirmed in the special case of the McVittie-like solution, where $\delta U$ corresponds to $U_{\rm static}$. Indeed inside the horizon, i.e. for $mr\ll 1$, we found
in \eq{NewtUrmsmall} that 
$U_{\rm static}=r_S/r$, while, again for $mr\ll 1$, the potential $\Phi$ is given by $r_S/(2 r)$,
see \eq{PhicPhifinal}. Thus, the results of the two approaches are fully consistent. 

The result 
$\delta U=2 \Phi$ is also consistent with that found using linear cosmological perturbation theory for the RR model. Indeed, in that case, for modes well inside the horizon, in the linearization   approximation one finds  $\delta U   = 2  (\Psi + 2 \Phi) $ [see  eq.~(4.10) of \cite{Dirian:2014ara}, neglecting the time derivatives as appropriate to the quasi-static approximation and retaining only the leading term in the
the limit $\hat{k}\equiv k/(aH) \gg 1$].
Together with $\Psi=-\Phi$ (which is due to the fact that in the RR model anisotropic stresses are negligible, see again \cite{Dirian:2014ara}), this gives again $\delta U=2 \Phi$. Once again, the advantage of the present discussion, with respect to the linearized treatment of \cite{Dirian:2014ara}, is that here we have not assumed that the density perturbations are small, but only that $|\Phi|$ is small, a condition that, in the absence of strong backreaction effects, includes the situation in which there is a full hierarchy of structures between the solar system scale and the cosmological scales.

To sum up, we have found a consistent solution of the form (\ref{pertans1}), (\ref{pertans2}), where the perturbations $\delta U$ and $\delta S$ induced by the perturbation of FRW (and therefore by non-linear structures, as long as one accepts that there is no large backreaction and non-linear structures are well described by a perturbed FRW metric) remain small. As a consequence, 
the solutions for $U$ and  $S$ retain their time dependence down to short scales, and  even on LLR scales we have  $G_{\rm eff} \simeq G/[1-(1/3)m^2 S_c(t)] $.

Finally, one might ask what happens if one tries to  force an exactly  static solution at the Earth-Moon scale. In GR this can be done, and the classic example  is the Einstein-Straus construction, in which one has again a central mass $M$ embedded in FRW, as in the McVittie case, but, rather than finding a metric that interpolates smoothly between the two, one just matches the  two solutions on a boundary surface $\Sigma$ (see e.g. \cite{Carrera:2008pi,Laarakkers:2001js} for review).
In the interior the metric is taken to be exactly the static \Sch metric generated by the mass $M$, while in the exterior  it is given by a FRW solution with energy density $\rho$. The two metrics are then matched by requiring that the induced metric on $\Sigma$ agrees on the two sides. This fixes the
matching radius $r_0$, that, with respect to the \Sch coordinates of the interior (that correspond to the physical FRW coordinate of the exterior region) turns out to be given by
$M=(4/3)\pi r_0^3\rho$, where $\rho$ is the energy density in FRW. 
In this case the solution for the metric is exactly static in the interior region. Nevertheless, if one studies the propagation of a scalar field obeying the equation $\Box\phi=0$ in this metric, one finds that the solution for the field in the inner region is time dependent~\cite{Laarakkers:2001js}. Technically, this comes from the fact that we must impose a matching condition for the scalar field at the surface $\Sigma$, and in this way the field inherits a time dependence even in the inner region. Similarly, in the RR model, because of the matching at $\Sigma$, the fields $U$ and $S$ will depend on time in the interior, thereby inducing a time-dependence in the Newton constant. This means that, in the RR model, there will be no exact Einstein-Straus solution, since even in the interior there will be a time-dependence in the metric induced by $G_{\rm eff}(t)$.

\section{No perfect screening for free}\label{sect:noscreeningDW}

We next examine how the Deser-Woodard model behaves with respect to LLR.  In ref.~\cite{Deser:2013uya}  a mechanism (referred as `perfect screening for free' in \cite{Woodard:2014iga}) was proposed, based on the observation that the scalar d'Alembertian $\Box$ has different signs when acting on functions of time and on functions of space, as it is clear from the flat-space limit $\Box\ra -\pa_t^2+\n^2$. As we saw in \eq{GammaDW}, the DW model can be written in terms of a function $f(X)$, where 
\be\label{XBoxR}
X\equiv \frac{1}{\Box} R\, .
\ee
(Note that, in the notations that we use for the RR model, $X=-U$; in the following we will use both $X$ and $U$). We first observe that $R$ is positive both in a cosmological setting, where the main contribution to the energy-momentum tensor comes from the cosmological fluid,  and in the local regions (such as the solar system, or the Earth-Moon system), where the main contribution comes from the mass of the non-linear structures, because in any case it is  dominated by the positive energy density $\rho$ [see e.g. \eq{menoRfromtrace} for the full expression in the RR model]. Then, in  ref.~\cite{Deser:2013uya,Woodard:2014iga} it is argued that in cosmology (where the time derivative dominates) $X\simeq \frac{1}{-\pa_t^2}R$ is negative   because of the minus sign in  $-\pa_t^2$, while it is positive in the regime dominated by structure formation, where the spatial derivatives dominate and $X\simeq \frac{1}{\n^2}R$.
Deser and Woodard then complete the definition of their model by stating that the function $f(X)$ is given by
\eq{f(X)} for $X<0$, and by  $f(X)=0$ for $X>0$.  According to the authors, this would automatically set to zero any effect of the non-local term in the solar system, and should allow the model to pass the solar system (and LLR) tests. Ref.~\cite{Woodard:2014iga} then discusses why, in the author's view, such a choice of $f(X)$ is not an unnatural  fine tuning.

First of all we observe that, if this mechanism actually worked, we could extend it trivially to the RR model. Indeed, upon integration by parts, $m^2R\Box^{-2}R$ can be written as $m^2 (\iBox R)^2$. We could then define the RR model writing $m^2 f(X)$, with $f(X)=X^2$ for $X<0$ and   $f(X)=0$ for $X>0$. 

However, this mechanism does not work, neither for the RR model nor for the DW model. The first reason is that  $1/\Box$ is not an algebraic quantity but an integral operator, and one cannot determine the sign of $X$ as done above. \Eq{XBoxR} is equivalent to the differential equation $\Box X=R$.  In a regime where the time derivative dominates this is equivalent to $-\pa_t^2X=R$, i.e., in our notation,  $\pa_t^2U=R$ [in flat space; the corresponding equation in FRW is  \eq{syU}]. Note that the equation for $U$ is the same in the DW and RR models, since in both cases it simply follows from the definition $\Box U=-R$. So, first of all, the fact that $R$ is positive only  provides  a statement about the sign of the second derivative of $U$ (in flat space; or of the corresponding differential operator in FRW), and not directly about $U$ itself. To integrate $\pa_t^2U=R$, or \eq{syU}, we also need the initial conditions on $U$. In the DW model one sets $U(t_i)=\dot{U}(t_i)=0$ at an initial time $t_i$,  chosen deep in RD (see e.g. eq.~(37) of \cite{Woodard:2014iga}). In this case, the integration of the differential equation indeed confirms that $U>0$, i.e. $X<0$. 

Consider now the regime dominated by spatial derivatives. In that case, in the flat-space limit, $\n^2X\simeq R$, i.e. $\n^2U\simeq -R$. The argument of ref.~\cite{Deser:2013uya,Woodard:2014iga} amounts to saying that, since $R$ is positive, the solution $X$ of $\n^2X= R$ is also positive. In general, this is  false. For instance,
\be
\n^2\(\frac{1}{r}\)=-4\pi\delta^{(3)}(\vx)\, ,
\ee
which shows that the positive function $1/r$ has a Laplacian given by a Dirac delta with {\em negative} coefficient. In other words, the Green's function of the Laplacian is 
\be\label{Green}
G(\vx-\vx')=-\frac{1}{4\pi |\vx-\vx'|}\, ,
\ee
with a crucial minus sign.
At short distances, both in the RR and DW models,  the dominant contribution to the Ricci scalar $R$ comes from the energy density $\rho$ and can be determined by taking  the trace of the Einstein equations, which  gives
$R\simeq 8\pi G\rho$.\footnote{The fact that $R$ at short scales reduces to the GR expression $8\pi G\rho$ is necessary in any model that aims at modifying GR at the horizon scale without spoiling solar system and laboratory tests, unless there is a vDVZ discontinuity and a corresponding Vainshtein mechanism.}
This source term is therefore localized in space (e.g. it can be approximated with a Dirac delta for a single mass such as the Earth, at $r\gg r_S$). In this case, the  solution of $\n^2X= 8\pi G\rho$ is 
\be
X(\vx)=-8\pi G\int d^3x' \, \frac{\rho(\vx')}{4\pi |\vx-\vx'|}\, .
\ee
A positive $\rho$ therefore generates a {\em negative} $X$, i.e.~a positive $U$, just as in the regime dominated by the time derivative.\footnote{Observe that the requirement that $\rho$ has compact support is important because it ensures the convergence of the convolution with the Green's function. The Ricci scalar also contains non-compact terms that depend on the auxiliary fields themselves (and that are specific to the theory considered, DW or RR; see  \eq{menoRfromtrace} for the expression in the RR model). Their contribution can be included iteratively, using first the zero-th order result for the auxiliary fields obtained including in $R$ only the dominant contribution coming from $\rho$, and using them to compute the source term $R$ to the next iterative order. Already to the first iterative order the source term $R$  no longer has compact support, and the solution cannot be obtained with the simple Green's function (\ref{Green}). There are however other techniques to deal with the problem. Technically, the issue  becomes very similar to the one encountered when one computes the gravitational-wave production from a localized system, such as a compact binary system, using  a post-Newtonian/post-Minkowskian expansion: in that case, the gravitational waves produced at leading order act themselves as non-compact sources for  gravitational-wave production at the next  order (see e.g. section~5.3 of \cite{Maggiore:1900zz} for  a pedagogical review). In any case, the point is that these corrections are calculable and finite. At short scales, where the dynamics is dominated by the matter density $\rho$, they are just a small correction to the leading term and they cannot affect the sign of the solution for $U$ computed to leading order. On cosmological scales the Ricci scalar $R$ is no longer dominated by $\rho$, but rather by the contribution of the auxiliary fields that are responsible for the effective dark energy, and $U$ is no longer necessarily positive. Indeed, in the RR model it oscillates as $\cos (mr)$, see \eq{NewtU}.
\label{foot:signLaplacian}
} 
In terms of $U=-X$,
\be
U(\vx)=2G\int d^3x' \, \frac{\rho(\vx')}{|\vx-\vx'|}\, .
\ee
Taking for instance $\rho(\vx)=M\d^{(3)}({\bf x})$, we get $U(r)=r_S/r$, as we already found in \eq{NewtUrmsmall}. In conclusion, the static solution for the auxiliary field  has the same sign as the cosmological solution, so that $U$ is always positive (i.e. $X$ is always negative), contrary to what is needed in the screening mechanism proposed in \cite{Deser:2013uya,Woodard:2014iga}.

Actually there is a second, more subtle, reason why this `perfect screening for free' mechanism does not work. As we  have seen in sections~\ref{sect:McV} and \ref{sect:pertFRW}, once we look for a solution 
valid across the whole range of scales, from the short scales relevant for LLR and solar system experiments, up to the cosmological scales, we find that the solution at short scales is actually well approximated by the sum of the static solution and the cosmological solution, $U(t,r)\simeq U_{\rm cosmo}(t)+U_{\rm static}(r)$. Note, in particular,  that the analysis of section~\ref{sect:pertFRW} based on the perturbed FRW metric can be repeated with basically no modification for the DW model. Indeed the equation to be studied, $\Box U=-R$, is the same for the RR and DW model, and in both cases one just needs to assume the validity of a perturbed FRW metric. What changes between the two models is the explicit expression of the Ricci scalar in terms of the auxiliary fields of the theory. However, this just means that $ U_{\rm cosmo}(t)$ will be the respective cosmological solution in each model.\footnote{Furthermore, in the DW model, on cosmological scales the explicit expression of $R$ as a function of time is the same as in $\Lambda$CDM because, by construction, the DW model has the same background evolution as  $\Lambda$CDM, while in the RR model   the evolution turns out to differ from that in $\Lambda$CDM only at the level of about $10\%$, so in practice, on cosmological scales, we are integrating the equation $\Box U=-R$ with a very similar source $R(t)$ while, at short scales, in both the DW and RR model $R$ reduces to the GR expression.} 
Now, numerically, for the Earth-Moon system $U_{\rm static}(r)\simeq r_S/r$ where $r_S$ is the \Sch radius of the Earth, about $1~{\rm cm}$, while $r$ is the Earth-Moon distance. Thus, $ r_S/r\sim 10^{-11}$ is a minuscule quantity, while $ U_{\rm cosmo}(t)={\cal O}(1)$ so, numerically, the solution for $U$ even on the scale of the Earth-Moon system is completely dominated by the cosmological solution.

In conclusion, first of all it is not true that, in the static solution, $U$ is negative; the static solution has the same sign as the cosmological solution. And, furthermore, the sign of the static solution  is irrelevant, because even at short scales $U$ is dominated by the cosmological solution. This means that the screening mechanism suggested in  \cite{Deser:2013uya,Woodard:2014iga}, based on setting $f(X)=0$ for $X>0$, does not work: $X=-U$ is always negative, both at cosmological scales and at short scales. As a consequence, the time-dependence of the effective Newton constant obtained on cosmological scales holds down to the LLR scale. In the DW model the latter is given by \eq{GeffDW}. This quantity has been computed numerically in \cite{Nersisyan:2017mgj,Park:2017zls} and one finds that, today, 
\be\label{GeffsuGDW}
\frac{1}{G_{\rm eff}}\frac{dG_{\rm eff}}{dz}={\cal O}(1)\, ,
\ee
where $z$ is the redshift, see  Fig.~4 of \cite{Park:2017zls}. In terms of the derivative with respect to cosmic time this means that, in the present cosmological epoch, 
\be\label{GeffsuGDWdt}
\frac{1}{G_{\rm eff}}\frac{dG_{\rm eff}}{dt}\simeq H_0\, ,
\ee
with a proportionality coefficient of order one. Thus, the DW model violates the LLR bound (\ref{dotGsuGH0}) by about three orders of magnitude and, just as the RR model,  is ruled out.

\section{Conclusions}\label{sect:conclusions}

In this paper we have elaborated on the suggestion in ref.~\cite{Barreira:2014kra}, that the limit on the time variation of Newton's constant from Lunar Laser Ranging could be in conflict with the RR model. The main question was whether one could extrapolate the result for $G_{\rm eff}(t)$  found on cosmological scales down to the scales relevant for Lunar Laser Ranging. The analysis that we have presented shows that this is indeed the case, unless some strong backreaction effect from nonlinear structures comes into play and provides a screening mechanism, a hypothesis that seems unlikely in view of much recent work on the backreaction problem. We have found that this indeed rules out the RR model [unless the parameter $u_0$ is very large $u_0\, \gsim \, {\cal O}(2000)$, which seems quite unnatural even from the point of  view of an earlier inflationary phase]. However the RT model, defined by \eq{RT}, passes this test because in this case
$G_{\rm eff}$ reduces to $G$ on small scales, see \eq{GeffGRTlargek}. We have further examined the situation in the Deser-Woodard model and we found that it suffers from the same problem as the RR model. The screening mechanism that was proposed for this model, based on the idea that $\iBox R$ changes sign from the cosmological regime to the short-scale regime, turns out to be technically incorrect, and even the 
Deser-Woodard model, with the form of the distortion function chosen so to mimic $\Lambda$CDM, is ruled out by Lunar Laser Ranging (at least, unless it can be supplemented by a different screening mechanism). 

As we briefly reviewed in the Introduction (see \cite{Belgacem:2017cqo} for more details), in the last decades have been proposed several variants of the idea of modifying phenomenologically the quantum effective action of gravity, through the addition of non-local terms  relevant in the infrared. At present, the RT model appears to be the only viable model that has passed all the tests performed, namely: (1) At the level of background cosmological evolution  it provides a viable FRW solution, with the nonlocal term acting as an effective dark energy, thereby providing accelerated expansion without a cosmological constant. (2) Cosmological perturbations are well behaved, both in the scalar and in the tensor sector. (3) Implementing the model in a Boltzmann code and  comparing with observations, one finds that  the model  fits  CMB, BAO, SNe, structure formation data and local $H_0$ measurements at a level statistically equivalent to $\Lambda$CDM.  (4) Gravitational waves propagate at the speed of light, complying with the limit from GW170817/GRB 170817A. (5) Solar system tests, and in particular the limit on the time variation of Newton's constant from  Lunar Laser Ranging, are satisfied.\footnote{Note also that there is  no problem with ghosts nor extra degrees of freedom, once one properly interprets these nonlocal models as derived from a quantum effective action. See in particular section~2.4 of \cite{Belgacem:2017cqo}, or sections 5-6 of \cite{Maggiore:2016gpx}, for detailed discussions.}

The fact that these requirements are so stringent, to the extent that they appear to select a single model, is quite interesting, and gives potentially important hints for attempts at deriving these  nonlocal terms from a fundamental theory, a step that is eventually essential to put these ideas on firmer grounds.

\vspace{5mm}\noindent
{\bf Acknowledgments.} We thank Alex Barreira and Baojiu Li for very useful discussions, already several years ago, that eventually led us to fully understand the importance of their observation on LLR. We also thank Giulia Cusin, Yves Dirian and Stefano Foffa for useful discussions and comments on the manuscript.
The work  of E.B.,  A.Fi. and M.M. is supported by the Fonds National Suisse and  by the SwissMap National Center for Competence in Research. The work of A.Fr. was supported by a Swiss
Government Excellence Scholarship and is currently supported from ERC Advanced Grant GravBHs-692951 and MEC grant FPA2016-76005-C2-2-P.

\appendix

\section{Static solution for the RR model}\label{App:Sch}

In this appendix we give a more detailed derivation of the results presented in section~\ref{sect:staticRR}.
To find the solution we follow the strategy used in \cite{Kehagias:2014sda} for the RT model: we first compute the solution in the limit  $mr\ll 1$, obtaining the leading term (and, for the metric, the next-to-leading term) of the solution in the regime $r\ll m^{-1}$; we then compute it in a Newtonian expansion around flat space, which gives a solution valid for $r\gg r_S$, and we finally match the two solutions in the overlap region $r_S\ll r\ll m^{-1}$, which will allow us to fix all coefficients of the solutions.

\subsection{Solution for $mr\ll 1$}

Let us begin with the expansion in powers of $mr$. To zero-th order, \eqs{eqab1}{eqab2}
reduce to their standard GR form, whose solution is given by
\be\label{solAB}
e^{-2\beta(r)}=1-\frac{r_S}{r}\,\, ,\qquad 
\a(r)=-\b(r)\, .
\ee
Plugging these expressions into \eq{eqU} we get
\be\label{UVgen}
U(r)=u_0-u_1\ln\(1-\frac{r_S}{r}\)\, , 
\ee 
so, at $r\gg r_S$,
\be\label{UVgenlarger}
U(r)\simeq u_0+u_1\frac{r_S}{r}\, .
\ee 
Until now, everything is just the same  as for the RT model. Plugging these expressions in \eq{eqS} we get an equation for $S(r)$ whose  solution is
\begin{eqnarray}\nonumber
&&\frac{S(r)}{r_S^2}=c_0-c_1\log\left({1-\frac{r_S}{r}}\right)-\frac{u_0}{3}
\left[\frac{1}{2}\(\frac{r}{r_S}\)^2+\frac{r}{r_S}+\log\left(\frac{r}{r_S}-1\right)\right]\\
&&-u_1\left[\frac{2}{3}\polylog\left(1-\frac{r_S}{r}\right)+\log\left(\frac{r}{r_S}-1\right)+\frac{1}{3}\frac{r}{r_S}-\frac{1}{3}\log\left({1-\frac{r_S}{r}}\right)\left(\frac{r}{r_S}+\log\frac{r}{r_S}\right)
\right. \nn\\
&&\hspace{8mm}\left.
-\frac{1}{6}\frac{r^2}{r_S^2}\log\left({1-\frac{r_S}{r}}\right)\right]\,,\label{Sfull}
\end{eqnarray}
where $\polylog (z)$ denotes the polylogarithm of order 2.
For $|z|<1$ (i.e. $r>r_S/2$ since, in our case, $z=1-\frac{r_S}{r}$) it is given by
\begin{equation}
\polylog(z)=\sum\limits_{n=1}^\infty \frac{z^n}{n^2} \, .
\ee
Expanding  for $r\gg r_S$, \eq{Sfull} reduces to
\be\label{Sofrlimit}
\frac{1}{r_S^2}S(r)\simeq  -\frac{u_0}{6} \(\frac{r}{r_S}\)^2- \frac{2u_0+3u_1}{6}\, \(\frac{r}{r_S}\) 
+{\cal O}\( \log \frac{r}{r_S} \)\, .
\ee
We can now plug these expressions for $U$ and $S$ into  \eqs{eqab1}{eqab2} to get the  corrections 
${\cal O}(m^2r^2)$
to the metric. Again, we limit ourselves to $r\gg r_S$. The result is
\be\label{solalphamr}
\alpha(r)\simeq -\beta(r)+\frac{u_0}{36} (mr)^2-\frac{u_0u_1}{18} m^2rr_S
+{\cal O}\(m^2r_S^2\log \frac{r}{r_S}\)\, ,
\ee
and
\be\label{solbetamr}
e^{-2\beta(r)}
=1-\frac{r_S}{r}\[ 
1+\frac{m^2 r^2}{36}
\(  c_1+c_2\frac{r}{r_S}\)  \]
\, ,
\ee
where
\be \label{c1c2u0u1}
c_1=-u_0+2u_0u_1+6u_1\, ,\qquad c_2=u_0(4+u_0)\, .
\ee 
In terms of $A(r)=e^{2\alpha(r)}$ and $B(r)=e^{2\beta(r)}$, again to leading order in the limit $r\gg r_S$, we  get
\bees
A(r)&=& 1-\frac{r_S}{r}\left\{ 
1+\frac{m^2 r^2}{36}
\[  (c_1+4u_0u_1+2u_0)+(c_2-2u_0)\, \frac{r}{r_S}\]  \right\}\, ,\label{solArmr}\\
B(r)&=& 1+\frac{r_S}{r}\left\{ 
1+\frac{m^2 r^2}{36}
\[  (c_1+2c_2)+c_2\frac{r}{r_S}\]  \right\}\, .\label{solBrmr}
\ees

\subsection{Solution in the Newtonian limit}\label{sect:appNewtlim}

As in \cite{Kehagias:2014sda},
we next compute the static solution in a Newtonian expansion over flat space. The resulting solution will be valid for $r\gg r_S$, but with no restriction on $mr$. For scalar perturbations, in the Newtonian gauge, the perturbed metric can be written as
\be\label{ds2rN}
ds^2 =  -(1+2 \Psi) dt^2 + (1 + 2 \Phi) 
\[ dr_N^2 +r_N^2(d\theta^2+\sin^2\theta\,  d\phi^2)\]
\, ,
\ee
where we denote by $r_N$ the radial coordinate in the Newtonian gauge, while we reserve the notation $r$ for  the radial coordinate in \Sch coordinates used in the previous subsection.
Comparing with the linearization of  \eq{ds2}, which is 
\be\label{ds2ablin}
ds^2=-(1+2\alpha) dt^2+(1+2\beta)dr^2 +r^2(d\theta^2+\sin^2\theta\,  d\phi^2)\, ,
\ee
one obtains~ \cite{Kehagias:2014sda} $r=(1+\Phi)r_N$ and
\be\label{relaz}
\a=\Psi\, ,\qquad \b=-r\Phi'\, .
\ee
We also expand the auxiliary fields as
$U=\bar{U}+\d U(r)$, $S=\bar{S} +\d S(r)$. However,
since, when expanding over Minkowski space, the background values $\bar{U}=\bar{S}$ are equal to zero, we simply write the perturbations as
$U(r)$ and $S(r)$, keeping  in mind  that they are first-order quantities, just as $\Phi(r)$ and $\Psi(r)$. 
The linearization of \eq{BoxURR} gives
\be\label{UPhiPsi}
\n^2\(U-2\Psi-4\Phi\)=0\, ,
\ee
while the linearization of $\Box S=-U$ gives
\be\label{n2SU}
\n^2S=-U\, .
\ee
Note that, in these equations, the Laplacian is with respect to the Newtonian coordinate $r_N$. However,  since 
all quantities on the left and right-hand sides are already  of first-order in the perturbations, we can equivalently use $r$.
The linearization of the $(00)$ component of \eq{Gmn} gives
\be\label{lin00a}
-2\n^2\Phi=\frac{m^2}{3}U+8\pi G\rho\, ,
\ee
that, combined with \eq{n2SU},  becomes
\be\label{lin00b}
\n^2\(\Phi-\frac{m^2}{6}S\)=-4\pi G\rho\, .
\ee
Finally, from the linearization of the $(ij)$ component of \eq{Gmn} we get, again upon use of  \eq{n2SU},
\be
\(\d_{ij}\n^2-\pa_i\pa_j\)\(\Phi+\Psi-\frac{m^2}{3}S\)=0\, .
\ee
In particular, upon contraction with $\d_{ij}$, we get
\be\label{linijb}
\n^2\(\Phi+\Psi-\frac{m^2}{3}S\)=0\, .
\ee
We now observe that \eqsss{UPhiPsi}{n2SU}{lin00b}{linijb} are identical to the equations found in 
\cite{Kehagias:2014sda} in the context of the analysis for the RT model, with the field $S$, that in the RR model is defined by $S=-\iBox U$, playing the role of the field that was called $S$ in the context of the RT model, and defined by the fact that the spatial component $S_i$ of the auxiliary field $S_{\mu}$ can  be decomposed  as  $S_i=S_i^{\rm T}+\pa_i  S$, where  $\pa_iS_i^{\rm T}=0$. This is not surprising; indeed, it was already observed in \cite{Maggiore:2014sia}, using directly the nonlocal formulation, that the RR and RT model are identical when linearized over flat space. Here we have found the same result using the formulation in terms of auxiliary fields.

We can therefore now read the solution of these equations from the corresponding results  in 
\cite{Kehagias:2014sda}. Note that  one might be tempted to rewrite \eq{linijb} as $\Phi+\Psi-(m^2/3) S=0$, or \eq{UPhiPsi} as $U-2\Psi-4\Phi=0$. However,  these equations have been derived  in the limit $r\gg r_S$, where the linearization over flat space is appropriate, and  are not valid for generic $r$. If a function $f$ satisfies $\n^2 f=0$ over all  of space (and we further impose the boundary condition that $f$ vanishes at infinity), then $f=0$. However, from the fact that a function $f$ satisfies $\n^2f=0$ at large $r$ we cannot conclude that $f$ itself is zero, even if we impose that it vanishes at infinity:  we still  have a freedom in the solution due to the fact that any function that,  at large $r$, approaches the form $f(r)\simeq c_1r_S/r$ satisfies $\n^2f=0$ at large $r$. The correct procedure is to fix  constants such as $c_1$ by performing the matching with the solution found with the expansion in powers of $mr$. We quickly review below the correct procedure for  finding the solution,
referring the reader to \cite{Kehagias:2014sda} for more details.

First of all, to derive the solution for the auxiliary fields it is convenient to combine \eqss{n2SU}{lin00b}{linijb} into an equation involving $U$ only,
\be\label{nam2U}
(\n^2+m^2) U=-8\pi G\rho\, .
\ee
Note that this equation implies that, for a localized source, as in the \Sch case that we are studying here, $U$ must go to zero at large distances. From this we can already anticipate that, when we perform the matching with the solution (\ref{UVgenlarger}) in the overlap region $r_S\ll r\ll m^{-1}$, we will get $u_0=0$. 
For $\rho(\vx)=M \d^{(3)}(\vx)$ (or, more generally, at distances $r$ much larger than the size or the localized source) the most general solution of \eq{nam2U} is
\be\label{U4Glarge}
U(r)=\frac{r_S}{r}
\[\cos (mr) + \b \sin (mr)\]\, ,
\ee
with $\beta$ an arbitrary constant. Plugging this into \eqs{UPhiPsi}{lin00a} we get the solution for $\Phi$ and $\Psi$ (again, valid only in the linearization region  $r\gg r_S$),
\bees
\Phi(r)&=&\frac{r_S}{2r}\left\{c_{\Phi} +\frac{1}{3}\[\cos (mr)+\b\sin(m r)\]\right\}\, ,\label{PhicPhi}\\
\Psi(r)&=&\frac{r_S}{2r}\left\{c_{\Psi} +\frac{1}{3}\[\cos (mr)+\b\sin(m r)\]\right\}\, ,\label{PsicPsi}
\ees
where $c_{\Phi}$ and $c_{\Psi}$ are constants, for the moment arbitrary, that reflect the freedom of adding to the solution at large distances a term $\propto 1/r$, as discussed above. Note that we still require that $\Phi$ and $\Psi$ go to zero at infinity, as appropriate for the metric perturbations generated by a localized source. Using \eq{relaz} we can read the corresponding expression for $\a(r),\b(r)$, and therefore for $A(r)\simeq 1+2\a(r)$ and $B(r)\simeq 1+2\beta(r)$. The result is \bees
\hspace*{-5mm}A(r)&=&1+\frac{r_S}{r}\left\{c_{\Psi} +\frac{1}{3}\[\cos (mr)+\b\sin(m r)\]\right\}\, ,\label{AA}\\
\hspace*{-5mm}B(r)&=&1+\frac{r_S}{r}\left\{ c_{\Phi}+\frac{1}{3}\[\cos (mr)+\b\sin(m r)\]
+\frac{mr}{3}\[\sin (mr)-\b\cos(m r)\]\right\} .\label{BB}
\ees
Thus, the solution for the metric found with a Newtonian expansion around flat space, which is valid for $r\gg r_S$ but with no restriction on $mr$, has three undetermined parameters $\beta, c_{\Psi}, c_{\Phi}$. In contrast, we have seen that the solution found with an expansion in powers of $mr$, again assuming $r\gg r_S$, given by \eqs{solalphamr}{solbetamr} [or, equivalently, by \eq{solBrmr} for $B(r)$ and the corresponding expression for $A(r)$] has two free parameters $u_0,u_1$. Quite nicely, all these parameters can be fixed by comparing the two solutions in the overlap region $r_S\ll r\ll m^{-1}$, where they are both valid. Indeed, taking  $mr\ll 1$ in \eqs{AA}{BB}   we get
\bees
A(r)&=&1+\frac{r_S}{r}
\(c_{\Psi} +\frac{1}{3}+\b mr-\frac{m^2r^2}{6}\)\, ,\label{solArsmallmr}\\
B(r)&=&1+\frac{r_S}{r}
\(c_{\Phi} +\frac{1}{3}+\frac{m^2r^2}{6}\)\, .\label{solBrsmallmr}
\ees
Comparing \eqs{solArmr}{solBrmr} to \eqs{solArsmallmr}{solBrsmallmr} we get $c_{\Phi}=2/3$, 
$c_{\Psi}=-4/3$, $\beta=0$, 
$u_0=0$ and $u_1=1$. The fact that $u_0$ is forced by the matching to vanish has important consequences. Indeed, if the coefficient of the term  $m^2r^2 (r/r_S)$ in \eqs{solArmr}{solBrmr} would have been of order one, this term would have become of order one for $r\,\gsim\, r_c=(r_Sm^{-2})^{1/3}$. This scale is much smaller that $m^{-1}$ since, from the comparison with  cosmological data, the mass scale $m$ of the nonlocal model is   of order $H_0$ and therefore $r_S\ll m^{-1}$. Thus, the solution given by \eqs{solArmr}{solBrmr} would have entered a `strong coupling' regime, where the expansion in powers of $mr$ breaks down, already at a critical value of $r_c$ such that $mr_c\ll 1$. However, since $u_0$ is fixed by the matching procedure to the value $u_0=0$, the coefficient of this term vanishes, both in $A(r)$ and in $B(r)$, and the solution (\ref{solArmr}), (\ref{solBrmr}) is valid until  $mr$ approaches a value of order one.

For $A(r)$ and $B(r)$ we therefore end up with the solution already announced in \eqs{NewtA}{NewtB} while, for $\Phi$ and $\Psi$,  \eqs{PhicPhi}{PsicPsi} can be rewritten as
\bees
\Phi(r)&=&\frac{r_S}{2r}\[1-\frac{2}{3}\sin^2\(\frac{mr}{2}\)\]\, ,\label{PhicPhifinal}\\
\Psi(r)&=&-\frac{r_S}{2r}\[1+\frac{2}{3}\sin^2\(\frac{mr}{2}\)\]\, ,\label{PsicPsifinal}
\ees
while for $U(r)$ we  get
\be\label{U4Glargefinal}
U(r)=\frac{r_S}{r} \cos (mr)\, .
\ee
We can now compute the corresponding solution for $S(r)$ (that was not explicitly studied in \cite{Kehagias:2014sda,Maggiore:2014sia}, but that we need here because we are eventually interested in matching the static solution for $S(r)$ to its cosmological solution, for obtaining the effective Newton constant).
Using \eq{U4Glargefinal}, \eq{n2SU} becomes
\be\label{n2SUapp}
\n^2S=-\frac{r_S}{r} \cos (mr)\, ,
\ee
which, as usual, is valid only for $r\gg r_S$. This can be rewritten as
\be\label{Sflateq}
\frac{d^2}{dr^2}(rS)=-r_S \cos(mr)\, ,
\ee
whose solution is
\be\label{solm2S}
m^2 S(r)=s_0+\frac{r_S}{r}\[ s_1+\cos(mr)\]\, ,
\ee
with $s_0,s_1$ dimensionless integration constants. In the limit $mr\ll 1$ this becomes
\be\label{solSofr}
m^2S(r)\simeq s_0+(s_1+1)\frac{r_S}{r}-\frac{1}{2}m^2 r r_S\, .
\ee
From \eq{Sofrlimit}, with $u_0$ and $u_1$ now fixed to $u_0=0$ and $u_1=1$, we know however that the asymptotic behavior of $S(r)$ when $mr\ll 1$ must be $S(r)\simeq -(1/2)rr_S$. Comparison with \eq{solSofr} then shows that the  term $s_0+(s_1+1)r_S/r $, which is a zero-th order term with respect to the small parameter $mr$, is actually absent and this fixes
$s_0=0$ and $s_1=-1$. Thus, in the end
\be\label{solm2Sfinalapp}
m^2 S(r)=-\frac{r_S}{r}\[ 1-\cos(mr)\]\, .
\ee
This result can be checked observing that, through \eq{Gmn}, the spatial dependence of $S(r)$ induces a spatial dependence of Newton's constant, because of the term $m^2S\Gmn$ in $m^2K_{\mu\nu}$. 
However, the resulting modification of Newton's constant must be contained in the motion of a 
test particle in the \Sch metric (\ref{ds2}), (\ref{NewtA}), (\ref{NewtB}). The latter shows that any effect vanishes in the limit $mr\ra 0$. This means that, in \eq{solm2S}, we must have $s_0=0$ and $s_1=-1$ so that, in the limit $mr\ra 0$, $m^2S(r)$ vanishes.

Observe that we have found that the solution $S$ of $\n^2 S=-U$ is everywhere negative, including at $mr\ll 1$,  even if the solution for $U$, at $mr\ll 1$, is positive. This is not in contradiction with the analysis performed in section~\ref{sect:noscreeningDW}, where we found that, with $R$ positive, the solution for $U$ of $\n^2 U=-R$ is positive, a result that invalidated the screening mechanism that had been proposed for the DW model. The point, as discussed in footnote~\ref{foot:signLaplacian} on page~\pageref{foot:signLaplacian},  is that for $R$ the leading contribution is given by the source density $\rho$, which has compact support. The solution, at leading order, can then be found with the Green's function (\ref{Green}). In contrast, in $\n^2 S=-U$ the source term $U$ does not have compact support and the solution can no longer be obtained with a convolution with this Green's function. Rather, we have obtained it from the direct integration of \eq{n2SUapp} that, thanks to the $1/r$ behavior of the source term, takes the simple form (\ref{Sflateq}).

\section{Details of the computation in the McVittie metric}\label{app:McV}

In this appendix we give the details of the derivation of \eq{BoxUansatz}.
We start from \eq{BoxUMcV} and we use the conditions (\ref{condPhi1}) and 
(\ref{condPhi2}).
Then, in \eq{BoxUMcV}, to lowest order all the occurrences of $\Phi$ can be dropped. Indeed, in $(1-2\Phi-H^2r^2)$, the term $2\Phi$ is always subleading with respect to one, both in the region $r\ll r_c$ where the McVittie metric is dominated by the central mass, and in the  region $r\gg r_c$ where the metric approaches FRW. This is different to what happens to the $H^2r^2$ term, that is totally negligible with respect to one in the mass-dominated region $r\ll r_c$, but is not parametrically small with respect to one at the horizon scale. Similarly, we can neglect the $\Phi$ and $r\pa_r\Phi$ terms in the coefficients of $\pa_t^2U$ and $\pa_tU$ in \eq{BoxUMcV} [or $\pa_t^2S$ and $\pa_tS$ in \eq{BoxSMcV}]. A similar analysis can be done for the dimensionless term $[2r\pa_r\Phi+r^2(\dot{H}+2H^2)]$ in \eqs{BoxUMcV}{BoxSMcV}: even in the regime $r_S\ll r\ll r_c$, which is dominated by the central mass, the term $2r\pa_r\Phi$ can be neglected since it is much smaller than one and therefore, as  of $(1/r)\pa_r U$ in \eq{BoxUMcV},  is negligible compared to the term 
$(1-2\Phi-H^2r^2)(2/r)\pa_rU\simeq (2/r)\pa_rU$ [and similarly for the term
$(1/r)\pa_r S$ in \eq{BoxSMcV}].

Of course, this does not mean that we have lost all the dependence on the central mass $M$. The mass enters through the source term $R$, e.g. through a term proportional to $\rho(\vx)=M \d^{(3)}(\vx)$ if, for the purpose of studying the solution at $r\gg r_S$, we approximate the mass as point-like. The situation is precisely the same as in the Newtonian approximation discussed in App.~\ref{sect:appNewtlim} where we found that, to leading order in $r_S/r$, $U$ satisfies \eq{nam2U}, $(\n^2+m^2) U=-8\pi G\rho$, where on the left-hand side there is no occurrence of $M$ nor of the metric perturbation $\Phi$, but the mass enters through $\rho$ on the right-hand side. Dropping all occurrences of $\Phi$, \eq{BoxUMcV}  becomes
\be\label{BoxUMcVPhizero}
(1-H^2r^2)\(\pa_r^2U+\frac{2}{r}\pa_rU\)-r(\dot{H}+2H^2)\pa_rU
-2rH\pa_t\pa_rU-\pa_t^2U-3H\pa_tU\simeq -R\, .
\ee
In the RR model the Ricci scalar $R$ can be determined by taking the trace of \eq{Gmn},
which gives (\ref{Rfromtrace}), where the trace of the energy-momentum tensor $T$  is given by
\eq{Ttrace}. 

We  now  wish to test whether (at least with these approximations) \eq{BoxUMcVPhizero} is   solved by the ansatz
\be\label{ansatz1App}
U(t,r)\simeq U_{\rm cosmo}(t)+U_{\rm static}(r)\, ,\quad
S(t,r)\simeq S_{\rm cosmo}(t)+S_{\rm static}(r)\, ,
\ee
where $U_{\rm cosmo}(t)$ is the cosmological solution discussed in section~\ref{sect:cosmoRR} and
$U_{\rm static}(r)$ is the static \Sch solution of section~\ref{sect:staticRR} (a completely analogous analysis can be performed for the equation obtained from $\Box S=-U$).

The terms that would forbid a separation of variables of this sort are those in which there is a dependence both on $r$ and on $t$. Several of these terms have already disappeared when we dropped $\Phi(r)$. Among the remaining terms, consider for instance the term
\be\label{term1}
\[1-H^2(t)r^2\] \(\pa_r^2U_{\rm static}+\frac{2}{r}\pa_rU_{\rm static}\)\, ,
\ee
that comes out when we insert the ansatz (\ref{ansatz1App}) in the first term in \eq{BoxUMcVPhizero}. A priori it depends both on $t$ and $r$.
However,  in the regime $r_S\ll r\ll m^{-1}$ we have seen that $U_{\rm static}\simeq r_S/r$, and then the right-hand side of \eq{term1} vanishes. More generally, 
the right-hand side of \eq{term1} will be suppressed by a factor $r_S/r$. For instance, 
using the solution (\ref{NewtU}) valid in the actual metric for the RR model, we get
\be
\pa_r^2U_{\rm static}+\frac{2}{r}\pa_rU_{\rm static}=-\frac{r_S}{r}m^2\cos(mr)\, ,
\ee
so indeed in the cosmological regime this term is suppressed by a factor $r_S/r$.  At cosmological distances this  suppression factor is of the order of the ratio of the \Sch radius of the central mass (about 1~cm for the Earth) to the size of the horizon  of the Universe, so it is totally negligible. Therefore the term (\ref{term1}) is only relevant 
in the mass-dominated region $r\ll r_c$. However, in this regime the term $H^2(t)r^2$ is  completely negligible compared to one, so we can replace everywhere the term (\ref{term1})   by 
\be\label{term1b}
\pa_r^2U_{\rm static}+\frac{2}{r}\pa_rU_{\rm static}\, ,
\ee
that now depends only on $r$.
The same analysis holds for the term $r^2(\dot{H}+2H^2)(1/r)\pa_rU_{\rm static}$ in \eq{BoxUMcVPhizero}, that is negligible at cosmological scales because of 
$(1/r)\pa_rU_{\rm static}$, and also in the mass-dominated region because there it is a correction ${\cal O}(H^2r^2)$ to the term $(2/r)\pa_rU_{\rm static}$ in \eq{term1b}. Thus, the term $r(\dot{H}+2H^2)\pa_rU_{\rm static}$ can be dropped everywhere.
Finally, the term $\pa_t\pa_rU$ vanishes on the ansatz (\ref{ansatz1}).

Thus, on the ansatz (\ref{ansatz1App}), the left-hand side of \eq{BoxUMcVPhizero} can be replaced by
\be\label{BoxUMcVapprox}
-\(\pa_t^2\Uc+3H(t)\pa_t\Uc\)+\(\pa_r^2U_{\rm static}+\frac{2}{r}\pa_rU_{\rm static}\) \, ,
\ee
which leads to the left-hand side of \eq{BoxUansatz}.
We see that the left-hand side of the equation nicely separates into the $r$-dependent part of $\Box U$ that determines  the static solution and the $t$-dependent part relevant for the cosmological solution. 

We next study the source term $-R$ that, according to \eqs{Rfromtrace}{Ttrace}, on the ansatz 
(\ref{ansatz1App}), becomes
\bees\label{menoRfromtrace}
-R&=&8\pi G\, \frac{ -\rho_{\rm cosmo}(t)+3p_{\rm cosmo}(t)  }{1-(m^2/3)[S_{\rm cosmo}(t)+S_{\rm static}(r)]  }\nn\\
&&-8\pi G\, \frac{M\d^{(3)}({\bf r})  }{1-(m^2/3)[S_{\rm cosmo}(t)+S_{\rm static}(r)]  }\nn\\
&& -\frac{m^2}{3}\, \frac{  3U+U^2-\pam S\paM U}{1-(m^2/3)[S_{\rm cosmo}(t)+S_{\rm static}(r)]  }\, ,
\ees
where, in the last line, in $3U+U^2-\pam S\paM U$, of course $U=U_{\rm cosmo}(t)+U_{\rm static}(r)$ and
$S=S_{\rm cosmo}(t)+S_{\rm static}(r)$.

We now observe, from \eq{solm2Sfinal}, that $m^2S_{\rm static}(r)$ is always very small compared to one, as long as $r\gg r_S$. Indeed, at cosmological scales, where $mr$ is not small, it is anyhow suppressed by
a factor $r_S/r$ while, if $r_S\ll r\ll m^{-1}$, it is suppressed both by a factor $r_S/r$ and by a factor $(mr)^2$. Thus, we always have  $|m^2S_{\rm static}(r)|\ll 1$ (as usual, as long as $r\gg r_S$), and we can neglect it in \eq{menoRfromtrace}. Thus, in the denominator of the first and second term in \eq{menoRfromtrace}, we can replace $\{1-(m^2/3)[S_{\rm cosmo}(t)+S_{\rm static}(r)]\}$ by
$[1-(m^2/3)S_{\rm cosmo}(t)]$.

 This means the the first term in \eq{menoRfromtrace} is exactly the same that one get in the cosmological solution, when one sets $U=U_{\rm cosmo}(t)$, while the second term is the  one would get when one studies the static solution, setting $U=U_{\rm static}(r)$, {\em except} that now $G$ is replaced by 
 $G/[1-(1/3)m^2S_{\rm cosmo}(t)]$, something which is missed when considering only the static solution.
The  combination $G/[1-(1/3)m^2S_{\rm cosmo}(t)]$ is nothing but the effective Newton constant found in \eq{Geff1suk2}, so the second term in \eq{menoRfromtrace}
can be rewritten as $ -8\pi G_{\rm eff}(t) M\d^{(3)}({\bf r})$.

Finally, consider the last term in \eq{menoRfromtrace}. When evaluated on $U_{\rm cosmo}(t)$, $S_{\rm cosmo}(t)$ it gives the term that, in the cosmological solution, combines with the  energy-momentum tensor of the cosmological fluid, and provides a dynamical dark energy, as studied in section~\ref{sect:cosmoRR}. When evaluated on the purely static solution it rather gives the non-linear terms that contribute to the static \Sch solution studied in section~\ref{sect:staticRR}. However, when we evaluate it on the ansatz
$U=U_{\rm cosmo}(t)+U_{\rm static}(r)$, $S=S_{\rm cosmo}(t)+S_{\rm static}(r)$,  $G$ is replaced by $G_{\rm eff}(t)$. In principle,
there is also a mixed term coming from $U^2-\pam S\paM U$, given by
\be
2U_{\rm cosmo}(t)U_{\rm static}(r)-g^{rt}(\pa_rU_{\rm static}\pa_tS_{\rm cosmo}+
\pa_rS_{\rm static}\pa_tU_{\rm cosmo})\, .
\ee
This term, however, can be neglected. In fact, at $mr\ll 1$, it is suppressed because of the $m^2$ factor in front of it, while at cosmological scales it is suppressed by the factors $r_S/r$ coming from the static solutions. As a result, $-R$ reduces to the expression given in the
right-hand of \eq{BoxUansatz}.

\section{Details of the computation in the perturbed FRW metric}\label{app:pertFRW}

In this appendix we present the details of the computation of section~\ref{sect:pertFRW}.
We begin by discussing the equation $\square U = -R$.  
Using the ansatz (\ref{pertans1}), (\ref{pertans2}) and making use of the condition $|\Phi|\ll 1$,
from \eq{Boxffirstorder} we get
\be
\Box U \simeq - \ddot{U}_c - 3 H \dot{U}_c  - 4 \dot{\Phi} \dot{U}_c + a^{-2}   \nabla^2 \delta U\, .
\ee
As in section~\ref{sect:McV}, to get the Ricci scalar we take the trace of \eq{Gmn}  in the perturbed FLRW metric, which gives\footnote{We have neglected the pressure term and we have written $T=-\rho$, as appropriate for our problem where the sources are non-relativistic structures. The pressure term could be easily reinstated, but plays no role in the following.}
\bees
-\( 1-\frac{m^2}{3} S\)R &=& - 8 \pi G  \rho \\
&&+ m^2 \[-U - U^2/3 -\frac{1}{3}(1+2\Phi)\dot{U}\dot{S} +  \frac{1}{3a^2}(1-2\Phi)\vec{\nabla} U \cdot \vec{\nabla} S \]\,  .\nn
\ees
Dropping again the ${\cal O}(\Phi)$ corrections and using the perturbative ansatz (\ref{pertans1}), (\ref{pertans2}), we get
\bees
-\( 1-\frac{m^2}{3} S_c\)R &=&
 \left[1+\frac{m^2 \delta S}{3-m^2 S_c} \right] \left[- 8 \pi G  \rho  + m^2 \(-U_{c} - \frac{U_{c}^2}{3} -  \frac{\dot{U}_{c}\dot{S}_{c}}{3} +   \frac{\vec{\nabla} \delta U \cdot \vec{\nabla} \delta S }{3a^2}\)\right] \nn\\
 &&+ m^2\(-\delta U -  \frac{2}{3} U_c \delta U   -  \frac{\dot{S}_c \delta \dot{U} }{3} -  \frac{ \dot{U}_c \delta \dot{S}}{3}  \)\,   ,
\ees
where $\delta  \dot{U}\equiv \pa_t(\delta U)$, $\delta  \dot{S}\equiv \pa_t(\delta S)$. Note that the term
proportional to $\vec{\nabla} \delta U \cdot \vec{\nabla} \delta S$ is formally of second order in the perturbations, but for the moment we keep it because it could be enhanced by the spatial derivatives (even if we will show below that it is actually negligible).
Now we further assume a quasi-static approximation 
\be\label{quasiApp}
|\delta  \dot{U} |\ll a^{-1} |\vec{\nabla} \delta U|\, , \qquad 
|\delta  \dot{S}  | \ll a^{-1}|\vec{\nabla} \delta S|\, ,
\ee 
and we also assume 
\begin{align}\label{eq:smallDeltaSApp}
|m^2 \delta S |\ll |\delta U |\, ,
\end{align}
whose validity will be  checked  a posteriori.
Thus, the equation  $\Box U = - R$ becomes 
\bees\label{eqddotU1}
  - \ddot{U_c} - 3 H \dot{U_c}  - 4 \dot{\Phi} \dot{U}_c + a^{-2}   \nabla^2 \delta U 
  &\simeq& \frac{ - 8 \pi G  \rho  + m^2 (-U_{c} - \frac{U_{c}^2}{3} -  \frac{\dot{U}_{c}\dot{S}_{c}}{3}  +   \frac{\vec{\nabla} \delta U \cdot \vec{\nabla} \delta S }{3a^2}  )  }{1-m^2 S_c/ 3} \nn\\
&&+m^2 \frac{-\delta U -  \frac{2}{3} U_c \delta U   -  \frac{\dot{S}_c \delta \dot{U} }{3} -  \frac{ \dot{U}_c \delta \dot{S}}{3}} {1-m^2 S_c/ 3}\, .
\ees
We now observe that, using \eq{eq:smallDeltaS}, 
\be
m^2 \frac{|\vec{\nabla} \delta U \cdot \vec{\nabla} \delta S |}{3a^2} \,\ll\,  
 \frac{|\vec{\nabla} \delta U|^2 }{3a^2}\, .
\ee
On the other hand $|\vec{\nabla} \delta U|^2 /(3a^2)$ is much smaller than the term 
$ a^{-2}   \nabla^2 \delta U$ on the left-hand side of \eq{eqddotU1}, since it contains one more power of the small perturbation $\delta U$ and has the same number of spatial derivatives. So, the  term
$m^2 |\vec{\nabla} \delta U \cdot \vec{\nabla} \delta S |/(3a^2)$  on the right-hand side of \eq{eqddotU1} can be dropped. Similarly, using first the quasi-static approximation (\ref{quasiApp}) and then \eq{eq:smallDeltaSApp}, we can show that, for modes inside the horizon also the terms $\dot{S}_c \delta \dot{U}$ and $\dot{U}_c \delta \dot{S}$ can be dropped. As for the term $-\delta U -  \frac{2}{3} U_c \delta U$ on the right-hand side of 
\eq{eqddotU1}, for modes well inside the horizon today, this is negligible compared to the term $ a^{-2}   \nabla^2 \delta U$ on the left-hand side,  since $m$ is of order $H_0$.

Thus,  we finally get
\be\label{eqddotU2}
  - \ddot{U_c} - 3 H \dot{U_c}  - 4 \dot{\Phi} \dot{U}_c + a^{-2}   \nabla^2 \delta U \simeq \frac{ - 8 \pi G  \rho  + m^2 (-U_{c} - \frac{U_{c}^2}{3} -  \frac{\dot{U}_{c}\dot{S}_{c}}{3}   )  }{1-m^2 S_c/ 3} .
\ee 
For vanishing $\delta U $ and $\Phi$ this reduces to the background equation for $U_c$ sourced by the background density $\rho_b$, which we subtract to get 
\be\label{dUrhominusrhobApp}
-4 \dot{\Phi} \dot{U}_c + a^{-2}   \nabla^2 \delta U \simeq  - \frac{ 8 \pi G   }{1-m^2 S_c/ 3}  (\rho-\rho_b),
\ee
as claimed in \eq{dUrhominusrhob}.
 
We finally need 
to check the consistency of the assumption~(\ref{eq:smallDeltaSApp}). To this purpose we consider  the equation $\Box S=-U$, that, in the same approximations become 
  \begin{align}
- \ddot{S_c} - 3 H \dot{S_c}  - 4 \dot{\Phi} \dot{S}_c + a^{-2}   \nabla^2 \delta S = - U_c - \delta U\, .
\end{align} 
From this we subtract the background equation $-\ddot{S}_c - 3 H \dot{S}_c = - U_c$ to get 
  \begin{align}\label{na2SSc}
a^{-2}   \nabla^2 \delta S = - \delta U  + 4 \dot{\Phi} \dot{S}_c .
\end{align} 
The term  $\delta U$ here must be dropped, since we found that $\delta U \simeq 2 \Phi$ and we have already dropped the terms ${\cal O}(\Phi)$ from the wave operator.\footnote{In any case,  this term would source a value of $\delta S$ such that $(k/a)^2\delta S_k\sim \delta U_k$, i.e
$m^2\delta S_k\sim (m^2/k_{\rm phys})^2 \delta U_k$, where $k_{\rm phys}=k/a$. Since $m\sim H_0$, for modes well inside the horizon this would give $m^2\delta S_k\sim (m^2/k_{\rm phys})^2 \delta U_k\ll \delta U_k$.}
As for the term $4 \dot{\Phi} \dot{S}_c $, we can estimate its typical value as follows.
The time evolution of $\Phi$ is determined by the 
typical  time-scale $T$ of the evolving matter structures, $\dot{\Phi}\sim \Phi/T$, while the cosmological background solution evolves with a time-scale given by $H$,
$\dot{S}_c \sim HS_c$. In the recent cosmological epoch, when $H\sim H_0$, we have
\be
\dot{\Phi}\dot{S}_c\sim \frac{\Phi}{T}\, H_0 S=\frac{\Phi V}{H_0T}\, ,
\ee
where, as usual, $V(t)=H_0^2S(t)$, and today $V$ is of order one. The contribution of structures that evolve on a time-scale $T$ shorter than the Hubble time is therefore such that the term $4 \dot{\Phi} \dot{S}_c $ in
\eq{na2SSc} dominates over $\delta U$. Nevertheless, even this source term induces a value of $\delta S$ that satisfies \eq{eq:smallDeltaS}. Indeed, \eq{na2SSc} gives 
\begin{align}
k_{\rm phys}^2 \delta S \sim   \frac{\Phi}{H_0T}\, .
\end{align} 
Writing $T\sim \lambda/v=1/(k_{\rm phys} v)$, where $\lambda=1/k_{\rm phys}$ is the typical size of a structure and $v$ its velocity, and recalling that $\Phi\sim \delta U$ and $m\sim H_0$, we get
  \begin{align}
|m^2 \delta S_k| \sim \frac{H_0}{ k_{\rm phys}} v |\delta U_k| \ll  |\delta U_k|\, ,
\end{align} 
since both $H_0/ k_{\rm phys}\ll 1$ and $v\ll 1$.
At early times (approaching matter-radiation equality) $\delta S$ is even more suppressed because its sources, which are perturbations around the background, were even smaller.


\bibliographystyle{utphys}
\bibliography{myrefs_massive}

\end{document}